%% file: paper.tex
\documentclass[letterpaper]{mn2e}
\input{psfig}
\input{macros_wkb}
\def\lesssim{\mathrel{\hbox{\rlap{\hbox{\lower4pt\hbox{$\sim$}}}\hbox{$<$}}}}
\def\gtrsim{\mathrel{\hbox{\rlap{\hbox{\lower4pt\hbox{$\sim$}}}\hbox{$>$}}}}
\begin{document}


\title[Environmental Effects on Satellite Galaxies]{Environmental
  Effects on Satellite Galaxies: The Link Between
Concentration, Size and Colour Profile}

\author[S.M. Weinmann et al.]          
       {\parbox[t]{\textwidth}{
        Simone M. Weinmann,$^{1}$\thanks{E-mail:simone@MPA-Garching.MPG.DE}  
         Guinevere Kauffmann$^{1}$, Frank C. van den Bosch$^{2}$,
      \\Anna Pasquali$^{2}$, Daniel H. McIntosh$^{3, 5}$,
         Houjun Mo$^{3}$, Xiaohu Yang$^{4}$, Yicheng Guo$^{3}$}\\ 
        \vspace*{3pt} \\
 $^1$Max-Planck-Institut f\"ur Astrophysik, Karl-Schwarzschild-Str.1, Postfach 1317, 85741 Garching, Germany\\  
       $^2$Max-Planck-Institute for Astronomy, K\"onigstuhl 17,  
           D-69117 Heidelberg, Germany\\ 
	   $^3$Department of Astronomy, University of Massachusetts, Amherst, MA 01003-9305, USA\\
       $^4$Shanghai Astronomical Observatory; the Partner Group of
           MPA, Nandan Road 80, Shanghai 200030, China\\
$^5$Department of Physics, 5110 Rockhill Road, University of Missouri --
Kansas City, Kansas City, MO 64110, USA}


\date{}

\pubyear{2008}

\maketitle

\label{firstpage}


\begin{abstract}
Using  the  SDSS  DR4  group  catalogue  of Yang  et  al.  (2007),  we
investigate  sizes,  concentrations,   colour  gradients  and  surface
brightness  profiles of  central  and satellite  galaxies. We  compare
central  and satellite  galaxies at  fixed stellar  mass, in  order to
disentangle  environmental from stellar  mass dependencies.  Early and
late type  galaxies are defined  according to concentration.   We find
that at fixed stellar mass,  late type satellite galaxies have smaller
radii and  larger concentrations than  late type central  galaxies. No
such  differences are  found for  early-type galaxies.   We  have also
constructed surface brightness and colour profiles for the central and
satellite galaxies  in our sample.   We find that  late-type satellite
galaxies  have a  lower  surface brightness  and  redder colours  than
late-type  central galaxies.   We show  that all  observed differences
between satellite  and central galaxies  can be explained by  a simple
fading model, in  which the star formation in  the disk decreases over
timescales of  2-3 Gyr after  a galaxy becomes a  satellite. Processes
that  induce  strong   morphological  changes  (e.g.  harassment)  and
processes  that strip  the galaxy  of its  entire ISM  need not  to be
invoked in order to explain the environmental dependencies we find.

\end{abstract}




\section {Introduction}
\label{intro}

It is  well known  that the properties  of galaxies residing  in dense
regions are  different from galaxies  in the field. Group  and cluster
galaxies  have been found  to have  redder colours,  to be  of earlier
morphological type, and to have a smaller specific star formation rate
compared  to more isolated  galaxies (e.g.  Dressler 1980;  Postman \&
Geller 1984; Poggianti et al. 1999; Balogh et al. 2004a and many other
studies).   Various physical mechanisms  leading to  the environmental
dependencies of  galaxy properties have been suggested,  most of which
should only  act on  satellites galaxies (galaxies  which are  not the
most  massive, central  galaxy in  their group).  Satellites  may lose
their   extended  gas   halo   (a  process   called  `starvation'   or
`strangulation', Larson, Tinsley \&  Caldwell 1980; Balogh, Navarro \&
Morris 2000), or the very outer  part of their gas disk by interaction
with  the intragroup  medium  or the  gravitational  potential of  the
group. This affects the fuel for future star formation and is expected
to lead to a slow decline of star formation, on timescales of $\sim$ 2
Gyr. Alternatively,  ram-pressure may strip galaxies  of a significant
fraction  of their  cold interstellar  medium, which  is the  fuel for
current  star formation and  is much  harder to  remove (Gunn  \& Gott
1972;  Quilis,  Moore \&  Bower  2000). This  would  lead  to a  rapid
truncation of  star formation.  Neither of theses  processes, however,
are expected  to significantly alter the morphologies  or stellar mass
profiles of  satellite galaxies. These may be  disturbed by high-speed
encounters  with   other  satellites  (Moore   et  al.  1996)   or  by
interactions with the tidal field  of their parent halo.  On the other
hand,  the   distribution  of   galaxy  stellar  masses   varies  with
environment,  shifting to  higher  masses in  regions with  increasing
local  density  or decreasing  group-centric  radius  (e.g. Balogh  et
al. 2001;  Bamford et al. 2008; van  den Bosch et al.   2008b).  If no
binning in stellar mass is done,  this effect accounts for part of the
environmental  dependencies observed,  due to  the  strong correlation
between galaxy properties and stellar mass (Kauffmann et al. 2003).

Recently, it has been found  by various studies that the environmental
effects  on  the  specific  star  formation rates,  and  hence  galaxy
colours,  are  more  fundamental  than the  effects  on  morphological
indicators (Kauffmann et al. 2004; Blanton et al. 2005a; Ball, Loveday
\& Brunner  2008; van  den Bosch et  al. 2008a;  but see also  Park et
al. 2007). For example,  galaxies in different environments which have
the same  colours do not  show much difference in  their morphological
indicators,  while a  colour difference  persists between  galaxies in
different environments  which have a similar  morphology (e.g. Blanton
et al. 2005a).

Nevertheless, it is well established that the environment does have an
effect on morphological indicators.  Also, it has been been found that
galaxies in dense regions might have slightly decreased radii compared
to their  counterparts in  less dense regions  (Blanton et  al. 2005a;
Kauffmann et al.  2004; but see Park et al. 2007).   In this paper, we
will quantify environment to first order by discriminating central and
satellite galaxies, following  van den Bosch et al.  (2008a, b), which
is  well  motivated,  as  explained  above. We  will  also  study  how
satellite galaxy  properties are  affected by the  mass of  their host
halo, which is a  physically better motivated estimator of environment
than the  local density,  see discussion in  Weinmann et  al. (2006a).
For the first time, we  show clearly that late type satellite galaxies
are  smaller in  size than  late type  centrals. We  also  address two
important  outstanding  questions  regarding  the  connection  between
galaxy morphology and environment.

First,  we would  like to  understand in  more detail  how environment
affects morphological indicators of  galaxies.  It has been found that
satellite galaxies  are, on  average, slightly more  concentrated than
centrals at fixed stellar mass (van den Bosch et al. 2008a).  However,
is this because a relatively  small fraction of satellite galaxies are
transformed  into   elliptical  galaxies  by   dramatic  environmental
processes? Or does it mean  that more subtle processes weakly increase
the bulge-to-total  luminosity for a majority  of satellites?  Insight
into  these  questions  can   be  gained  by  investigating  the  full
distribution  in  morphological indicators  of  central and  satellite
galaxies.  This  is done in section \ref{morph}  for the concentration
index. We  present evidence that elliptical galaxies  are not produced
by environmental effects.

Second, we  investigate the origin of the  difference in morphological
indicators  between   central  and  satellite   galaxies.   Two  basic
scenarios  seem possible.  First,  environmental processes  may affect
the stellar  mass distribution  in satellite galaxies,  rendering them
more  compact.  This  could for  example occur  by  disk instabilities
induced by  the potential of the  group, tidal stripping  of the outer
stellar  disk  etc.   The  second  option is  that  the  stellar  mass
distribution is  not affected  and that it  is the  light distribution
which changes. If  the star formation rate in  a galaxy decreases, the
disk fades due to ageing  of the stellar population. Since this effect
is expected to be weaker in  the central, older regions of the galaxy,
this naturally leads to  an increase in the bulge-to-total luminosity.
To find  out which  possibility is more  likely, we study  the surface
brightness  profiles, colour  profiles  and stellar  mass profiles  in
section  \ref{profile}.  We  find  indications  that  the  differences
between central and satellite galaxies  can entirely be explained by a
decreased star  formation rate in satellites, which  affects the shape
of the surface brightness  profile. Using stellar population synthesis
models, we also  show that a slow decline in  the star formation rate,
like in the `starvation' scenario, fits our data best.

In section \ref{data} we present our data set and the associated group
catalogue, assess  the completeness  of our satellite  sample, explain
our weighting scheme and  discuss morphological indicators. In section
\ref{results} our  results are  presented.  In section  \ref{model} we
attempt to understand the  colour profiles of satellite galaxies using
a stellar  population synthesis model.   A discussion of  our results,
and  comparison  with  previous   studies,  is  presented  in  section
\ref{Discussion},  followed  by a  summary  in section  \ref{summary}.
Throughout this paper, all  magnitudes, stellar masses, sizes etc. are
scaled according to $H_{0}=70\  {\rm km}\ {\rm s}^{-1} {\rm Mpc}^{-1}$
and have  to be changed accordingly  when $H_{0}$ is  varied. The only
exception  are halo masses,  which are  given in  $h^{-1}M_{\odot}$ to
facilitate  comparisons  to  results  in  our  previous  papers.   The
abbreviation `log' refers to a 10-based logarithm.

\section{Data}
\label{data}
The  analysis in  this paper  is based  on the  SDSS DR4  galaxy group
catalogue of Yang et al.  (2007, hereafter Y07).  This group catalogue
has been constructed  by applying the halo-based group  finder of Yang
et al. (2005) to the  New York Value-Added Galaxy Catalogue (NYU-VAGC;
see  Blanton et  al.  2005b).  From this  catalogue, Y07  selected all
galaxies  in  the Main  Galaxy  Sample  with  an extinction  corrected
apparent  magnitude brighter  than $m_{r}=18$,  with redshifts  in the
range $0.01<z<0.20$  and with a  redshift completeness $C_{z}  > 0.7$.
Additionally,  7091  galaxies  have  been  included  which  have  SDSS
photometry but  with redshifts  taken from alternative  surveys, which
results in a total number  of galaxies included in the group catalogue
of  369,447.  We  use  Petrosian magnitudes  and  colours (Strauss  et
al. 2002), corrected for  galactic extinction (Schlegel, Finkbeiner \&
Davis 1998)  and k+e-corrected to $z=0.1$, using  the method described
in  Blanton et  al. (2003b).   Stellar masses  are computed  using the
relations between stellar mass-to-light ratio and colour given by Bell
et al.  (2003), following van den  Bosch et al.   (2008a). Halo masses
for each identified galaxy group  were calculated based on the ranking
in  the total  characteristic  stellar mass  (see  Y07 and  references
therein for details).  No halo masses have been  obtained for a subset
of 66,263 galaxies, however, these  galaxies are also included in what
follows.

We have checked whether employing an alternative stellar mass estimate
affects our results by repeating part of
our analysis  using the publicly  available stellar masses
for the SDSS DR4 obtained by Kauffmann et al. (2003), who use spectral
information rather than colours and luminosities to obtain the
stellar mass-to-light ratio.  No significant differences were found in
our results.

  We  use Petrosian  half-light radii  as given  in the  SDSS.  If not
  otherwise  specified, $R_{50}$  and  the terms  `size' and  `radius'
  refer to  the physical Petrosian half-light radius  in the $r$-band,
  in kpc.  Following Shen et al. (2003), we have excluded all galaxies
  with an  $r$-band surface  brightness $\mu_{\rm 50}<23.0$  mag ${\rm
    arcsec}^{-2}$ and all galaxies  with Petrosian half-light radii in
  the  $r$-band below  1.6  arcseconds from  our  sample.  The  latter
  criterion affects a significant  number of galaxies and is motivated
  by the  following two points.   First, as galaxy sizes  approach the
  size of the seeing disk (with a median seeing in the SDSS DR4 of 1.4
  arcseconds,  Adelman-McCarthy et  al. 2007),  flux  measurements are
  affected, as shown in Blanton  et al.  (2001).  Second, very compact
  galaxies might be misclassified  as stars, although this should only
  happen in rare cases (Strauss et al. 2002). Unlike Shen et al. 2003,
  we do  not exclude galaxies  with apparent magnitudes  brighter than
  $r=15$,  since  this only  affects  a  very  small fraction  of  the
  galaxies  in  our  sample.   We  have  checked  that  applying  this
  additional criterion  does not  lead to any  visible changes  in our
  figures. Another difference to the analysis by Shen et al. (2003) is
  that we restrict our analysis  to galaxies with stellar masses above
  log($M/M_{\odot}$)=9.75.  At  masses below  this  limit, the  actual
  distribution  of the  radii of  early type  galaxies is  expected to
  contain a  significant contribution  from very small  galaxies (with
  radii below 0.5  kpc), as visible in Fig. 10 in  Shen et al. (2003).
  These very small  galaxies only make it into our  sample if they are
  at redshifts below z=0.02,  where statistics is insufficient to give
  meaningful results. We also apply this cut to our late type galaxies
  to have a consistent lower stellar mass cut in the entire sample.

After applying all the selection  criteria listed above, our sample of
group  galaxies consists  of  275,890 galaxies,  of  which 47,433  are
satellites. This means  that most of our sample  galaxies are centrals
which do  not have satellites that  are bright enough  to be detected.
Of  the satellites  galaxies, approximately  1/4 resides  in clusters,
i.e. groups with dark matter masses above $10^{14} h^{-1}M_{\odot}$.

\subsection{Completeness of Satellite Sample}
\label{complete}
In  order   to  check  the  completeness  and   contamination  of  our
satellite/central samples, we  make use of a group  catalogue that has
been built from a mock  galaxy redshift survey (MGRS) which mimics the
SDSS DR4.   Only luminosities,  and not stellar  masses, are  used for
this test,  but we expect this  to have little influence  on the final
results. The MGRS  is constructed by populating dark  matter haloes in
numerical simulations of  cosmological volumes with galaxies according
to the conditional luminosity  function (CLF) as described in Cacciato
et al.  (2008) and Yang et  al. (2008). By construction,  the MGRS has
the  same  luminosity  function   and  clustering  as  a  function  of
luminosity as the  galaxies in the SDSS.  The same  MGRS has also been
used in Y07 to test the performance of the group finder; more detailed
results on  the contamination and completeness of  the group catalogue
we have used in our analysis can be found there.

Here, we check the completeness and contamination of our satellite and
central   samples.   Completeness  is   the  fraction   of  satellites
(resp. centrals) in the redshift survey which are correctly identified
as  such by  our group  catalogue.  Contamination is  the fraction  of
satellite (resp.  central) galaxies in our group  catalogue which were
falsely identified  as such and are centrals  (satellites) in reality.
We  find  that our  satellite  sample  is 87  \%  complete  and has  a
contamination of 23  \%. The sample of central  galaxies, on the other
hand,  is  95  \% complete  and  has  a  contamination  of 3  \%.  The
incompleteness of the satellite  sample is caused by misclassification
of satellites  as centrals,  either because the  entire group  has not
been identified  as such, or  because the group  finder underestimates
the size of a group and does not include satellites which are far away
from the center.  The contamination of the satellite  sample is caused
by  the inclusion  of (isolated)  central galaxies  as  group members,
which can happen if they reside in the vicinity of a group, especially
along the  line of  sight.  The incompleteness  of the  central galaxy
sample mainly originates from events where the group finder is merging
two separate groups into one.  The contamination of the central galaxy
sample  is caused  by the  group finder  occasionally breaking  up one
group into smaller units.  These issues are also discussed in Y07.

In  principle, it  would  be simple  to  correct the  average of  some
parameter  for  the satellite  or  central  sample for  contamination.
Incompleteness  is not an  issue here,  since we  can assume  that the
galaxies lost to our central  or satellite sample are not different in
their properties  than those which we have  identified correctly.  Let
us assume  that we have  obtained some average  quantity $\hat{p}_{\rm
  cen}$ and  $\hat{p}_{\rm sat}$ in our  analysis, while $\bar{p}_{\rm
  cen}$   and   $\bar{p}_{\rm  sat}$   are   the   averages  for   the
(hypothetical)  uncontaminated  central  and satellite  sample.  Then,
given  a contamination  $c_{\rm sat}$  of the  satellite sample  and a
contamination $c_{\rm cen}$ of the central population,
\begin{equation}
\label{corr3}
\hat{p}_{\rm sat}=  (1-c_{\rm sat})  \cdot \bar{p}_{\rm sat}  + c_{\rm
  sat} \cdot \bar{p}_{\rm cen}
\end{equation}
\begin{equation}
\label{corr4}
\hat{p}_{\rm  cen}= c_{\rm  cen} \cdot  \bar{p}_{\rm sat}  + (1-c_{\rm
  cen}) \cdot \bar{p}_{\rm cen}.
\end{equation}
It is then straightforward to find that
\begin{equation}
\label{corr1}
\bar{p}_{\rm  sat}= (\hat{p}_{\rm  sat} -  \frac{c_{\rm sat}}{1-c_{\rm
    cen}}   \cdot  \hat{p}_{\rm  cen})   \cdot  \frac{1}{1   -  c_{\rm
    sat}-\frac{c_{\rm sat}c_{\rm cen}} {1-c_{\rm cen}}}
\end{equation}
and
\begin{equation}
\label{corr2}
\bar{p}_{\rm  cen}= (\hat{p}_{\rm  cen} -  \frac{c_{\rm cen}}{1-c_{\rm
    sat}}   \cdot  \hat{p}_{\rm  sat})   \cdot  \frac{1}{1   -  c_{\rm
    cen}-\frac{c_{\rm cen}c_{\rm sat}} {1-c_{\rm sat}}}
\end{equation}

Making the approximation  that the central sample is  100 \% complete,
we find  that $\bar{p}_{\rm  cen} - \bar{p}_{\rm  sat} \sim  1.3 \cdot
(\hat{p}_{\rm cen} - \hat{p}_{\rm sat}$), meaning that all differences
between  some average  quantity  for satellites  and  centrals can  be
expected to be higher by around 30  \% in reality than in what we will
find in our analysis.  It would be simple to correct for contamination
in  what follows. However,  the degree  of contamination  occurring in
reality  is  unfortunately  rather  uncertain, since  it  varies  with
stellar mass  and luminosity and has  also been found to  be depend on
the type  of MGRS  used. A  discussion of these  issues is  beyond the
scope of this paper.  We choose to present all results uncorrected for
contamination. However, we caution the  reader to bear in mind that we
likely underestimate the difference  between the properties of central
and satellite galaxies by around 30 \%.

\subsection{Treatment of Selection Effects}
\label{sec:selection}
In order to  model the selection function of the  survey, we give each
galaxy a weight which is  inversely proportional to the maximum volume
out to which  it can be observed.  This maximum  volume depends on the
absolute magnitude  of the galaxy,  its physical size and  its surface
brightness. We largely follow Shen et al. (2003) in our calculation of
the  galaxy weights.   First, we  calculate the  maximum  redshift and
luminosity  distance out  to  which a  galaxy  with apparent  $r$-band
magnitude  $r$  and an  effective  flux  limit  $r_{\rm max}$  can  be
observed:
\begin{equation}
d_{\rm L}(z_{\rm max, m})=d_{\rm L}(z)10^{-0.2(r-r_{\rm max})},
\end{equation}
where $d_{\rm L}$ is the  luminosity distance.  Next, we calculate the
maximum angular diameter  distance out to which a  galaxy with angular
Petrosian half-light radius $R_{50, \rm{ang}}$ can be resolved given a
lower limit on the angular size of 1.6 '':
\begin{equation}
d_{\rm A}(z_{\rm max,r})=d_{\rm A}(z)\frac{R_{50, \rm{ang}}}{1.6''}
\end{equation}
and $d_{\rm  A}={d_{\rm L}}/(1+z)^2$.   We also calculate  the maximum
redshift out  to which a galaxy  with surface brightness  $\mu$ can be
observed  given the  limit in  surface  brightness of  23.0 mag  ${\rm
  arcsec}^{-2}$   and  taking   into   account  cosmological   surface
brightness dimming proportional to $(1+z)^4$:
\begin{equation}
z_{\rm {max}, \mu}=(1+z) \times 10^{(23-\mu_{\rm 50})/10}-1
\end{equation}
We obtain the maximum redshift out to which a galaxy can be seen with

\begin{eqnarray*}
z_{\rm  max}={\rm Min}[z_{\rm  max, m},  z_{\rm max,  r},  z_{\rm max,
    \mu}, 0.2],
\end{eqnarray*}
where $z=0.2$ is the redshift limit of our sample, as mentioned above.
For a flat cosmology, the  maximum volume\footnote{Note that we do not
  use  the ``conditional''  maximum volume  like Shen  et  al. (2003),
  since  we  do   not  examine  the  dependence  of   galaxy  size  on
  luminosity.} out to which a galaxy can be observed is then
\begin{equation}
V_{\rm  max}=\frac{4}{3}  \pi \left[\frac{d_L(z_{\rm  max})}{(1+z_{\rm
      max})}\right]^3.
\end{equation}

\subsection{Quantitative Morphological Indicators}

One  of the  galaxy properties  we  intend to  study in  this work  is
morphology. Different indicators of  morphology have been used in past
studies to  classify galaxies in the  SDSS.  The ones  regarded as the
most reliable  are probably eyeball  classification and bulge-to-total
luminosity ($B/T$ hereafter), which  are not easily obtained for large
samples of galaxies (but see Lintott et al.  2008; Allen et al. 2006).
The  morphological  indicators which  are  most  widely  used are  the
S\'{e}rsic index  $n$ and  concentration $C$, both  publicly available
for  the  SDSS.   Below,  we  discuss  the  properties  of  those  two
morphological indicators and test their correlation with $B/T$.

\subsubsection{Comparing Concentration and S\'{e}rsic index}
\label{comp_morph}
In the SDSS, the concentration $C$ is defined as the radius containing
90 \% of the Petrosian flux  divided by the radius containing 50 \% of
the Petrosian flux of a galaxy.  As shown by Strateva (2001), Nakamura
et al. (2003)  and Shimasaku et al. (2001),  concentration can be used
to roughly distinguish  between early type and late  type galaxies, at
least for  the relative massive and  luminous galaxies we  use in this
study.  Note  that $C$ as given  in the SDSS is  not seeing corrected.
The S\'{e}rsic index $n$, on the other hand, is a measure of the shape
of the surface brightness profile $I(r)$, which can be fitted with
\begin{equation}
I(r)=I_{0} \cdot {\rm exp}\large[-(r/r_{0})^{1/n}\large]
\end{equation}
according  to  S\'{e}rsic  (1968).   Pure  disk  galaxies  with  their
exponential profile mostly fall into  the peak around $n$=1, while the
meaning  of $n$  becomes more  unclear  as it  increases.  Blanton  et
al. (2005a) applied a  1-dimensional S\'{e}rsic fit to the azimuthally
averaged  surface  brightness   profiles  of  the  NYU-VAGC  galaxies,
accounting for the effects  of seeing.  Thus, they obtained S\'{e}rsic
indices  $n$ and  also  derived S\'{e}rsic  half-light  radii for  the
galaxies   in  their  sample\footnote{The   Blanton  et   al.  (2003a)
  S\'{e}rsic  indices  and sizes  used  by  Shen  et al.  (2003)  were
  partially  incorrect   and  have   been  corrected  in   Blanton  et
  al. (2005a).}.

\begin{figure}
\centerline{\psfig{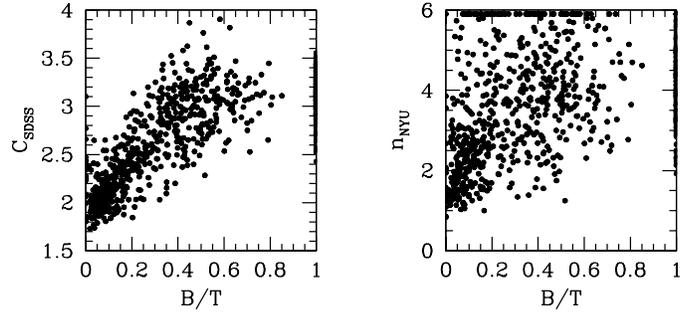}}
\caption{Concentration  $C$  (left   panel)  and  S\'ersic  index  $n$
  (according to  Blanton et al.  2005a, right panel),  plotted against
  the   bulge-to-total   ratio,  $B/T$,   as   obtained   in  the   2D
  multi-component fits of Gadotti (2008). $n$ and $C$
  are  both  given  in the  $r$-band,  while  $B/T$  is given  in  the
  $i$-band.  }
\label{fig:dim}
\end{figure}

In Fig.  \ref{fig:dim}, we show the concentration  parameter $C$ (left
panel) and the S\'ersic index according to Blanton et al. 2005a (right
panel), plotted against $B/T$ for  a sample of 852 face-on ($b/a \geq$
0.9), 0.02  $\leq z  \leq$ 0.07, morphologically  undisturbed, $M_{\rm
  star}  \geq 10^{10}$  galaxies.  Their $B/T$  has  been obtained  by
Gadotti (2008),  using   a  2-dimensional
bulge/disc/bar decomposition  and a  bulge profile with  variable $n$,
and  is seeing corrected.   Clearly, the  correlation between  $C$ and
$B/T$ is significantly better than  between $n$ and $B/T$. See Gadotti
(2008) for a more detailed analysis.
 
The poor correlation of $n$ with  $B/T$ indicates that $C$ is a better
proxy of galaxy  morphology. However, as mentioned before,  $C$ is not
corrected  for  seeing  effects.  To  test the  effect  of  seeing  on
concentration, we  create a subsample of galaxies  with physical sizes
between  3 and 4  kpc, with  masses above  log$(M/M_{\odot})$=10.8 and
with $\mu_{50}<22.63$  mag $\rm  arcsec^{-2}$. For this  subsample, we
are complete  up to $z=0.1$.  We  find that while  the average angular
sizes decreases from 4.5 times the seeing to only 1.4 times the seeing
between a redshift of 0.025 and 0.1, the average concentration changes
by less than 2 \%.  This  indicates that even when the angular size of
the galaxy  approaches the seeing, the average  concentration does not
seem to be strongly affected.  We have also checked the correlation of
the  seeing-corrected $C$ derived  from the  S\'{e}rsic fits  given by
Blanton et  al. (2005a) with $B/T$  and found that  the correlation is
similar to the one of $n$ with $B/T$.

Since the galaxies  in the sample shown in  Fig. \ref{fig:dim} are all
face-on, low-$z$ galaxies, we  cannot exclude the possibility that $n$
is superior to $C$ for  inclined and distant galaxies. However, we are
not aware of  any reason why this would be  the case, especially since
seeing seems to  have only a minor effect.  We  thus conclude that $C$
is likely to be a better  estimator of morphology than $n$ and we will
use $C$ throughout our paper.

\subsubsection{The Effects of Inclination on Morphological
Classification}
\label{inclination}
Both $n$ and $C$ are affected by inclination (e.g. Maller et al. 2008;
Bailin \&  Harris 2008). In  particular, late-type galaxies  which are
viewed edge-on often have values of  $C$ and $n$ typical of early type
galaxies.  For this reason, Maller et al. (2008) imposed an additional
criterion  on the  axis ratio,  $b/a$, classifying  all  galaxies with
$b/a<0.55$ as late  types.  The disadvantage of this  approach is that
edge-on S0  galaxies will be counted  as late types,  while face-on S0
galaxies will generally be counted  as early types. The same may apply
to other disk galaxies with a large bulge component.

Edge-on   disk  galaxies   are  significantly   reddened,   and  their
luminosities, sizes and stellar masses might be biased with respect to
galaxies viewed face-on  (e.g. Maller et al. 2008;  Shao et al. 2007;
Driver et al. 2007).
Therefore,  one could also  consider removing  all galaxies  with high
inclination from the sample. However, galaxies with $b/a<0.55$ make up
more than half of all the late type galaxies in our sample. Therefore,
our  standard approach will  be to  keep the  edge-on galaxies  in the
sample.  We will  also in  general not  use $b/a$  as a  criterion for
discriminating early  and late type  galaxies.  But we will  report on
how making these alternative cuts changes our results in what follows.

\begin{figure*}
\centerline{\psfig{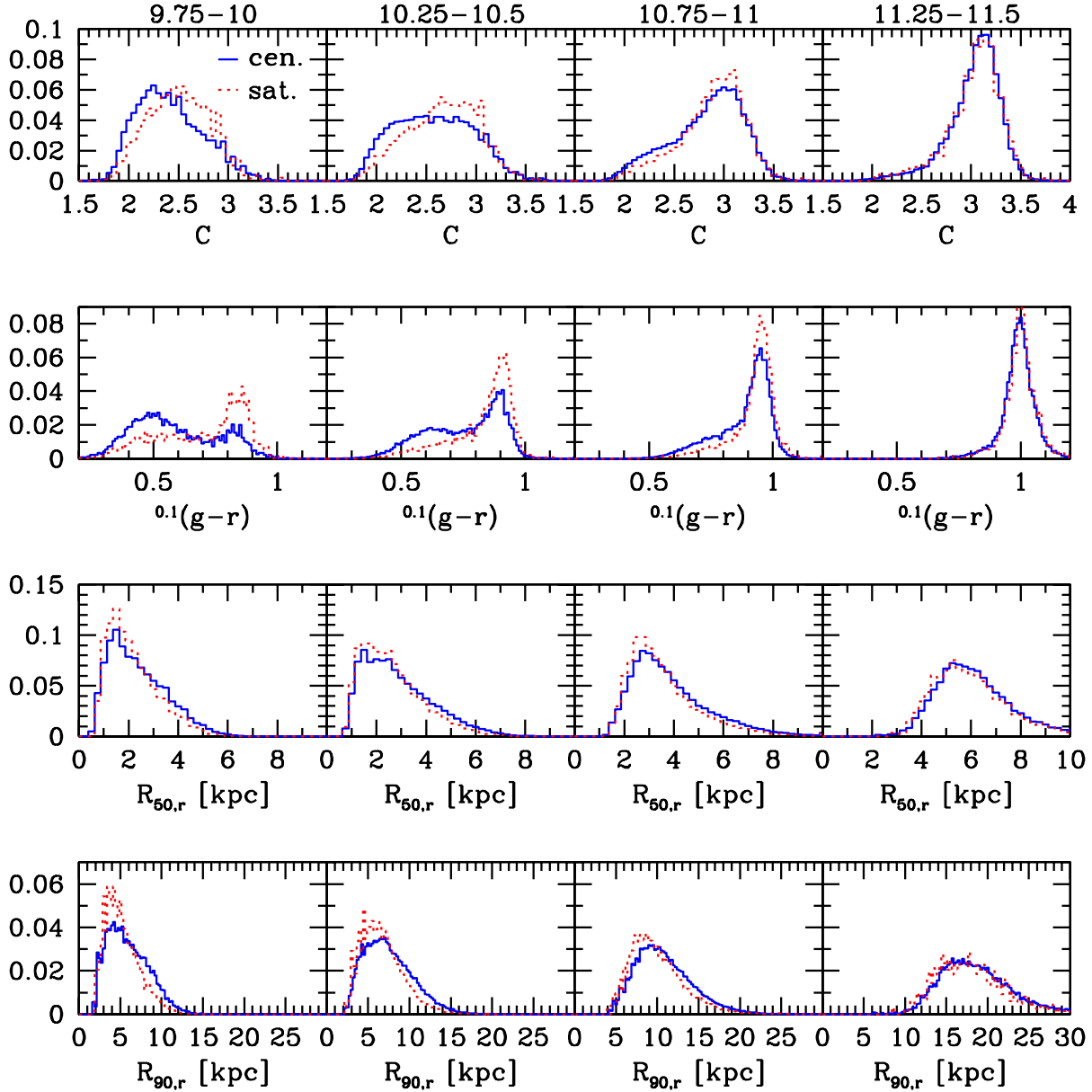}}
\caption{The  weighted distribution  of concentration  $C$  (top row),
  $^{0.1}(g-r)$ (second row from top), the physical Petrosian $r$-band
  half-light  radius $R_{50}$ (third  row from  top) and  the physical
  radius containing 90 \% of the Petrosian light $R_{90}$ (bottom row)
  in  four  different bins  in  log(stellar  mass),  as indicated,  for
  central  (solid  blue  line)  and  satellite  galaxies  (dotted  red
  line). Distributions have been normalized in each bin for satellites
  and centrals separately. Units on the y-axis are arbitrary.}
\label{fig:histo}
\end{figure*}

\section{Results}
\label{results}
In this  section, we compare  the properties of central  and satellite
galaxies in  our sample at  fixed stellar mass  in order to  probe the
influence    of   satellite-specific   environmental    processes   on
galaxies.  This comparison  is  based on  two  assumptions. First,  we
assume  that statistically  speaking, the  properties of  a  galaxy of
fixed  stellar mass  are  independent  of redshift,  at  least in  the
redshift   range  where   most   of  today's   satellites  have   been
accreted.  Second, we  assume  that a  galaxy  does not  significantly
change its  mass after becoming a satellite.  Under these assumptions,
comparing  satellite  galaxies of  today  to  equally massive  central
galaxies  of today is  equivalent to  comparing satellite  galaxies of
today to their progenitor central galaxy population at some (slightly)
higher redshift. See  van den Bosch et al. (2008a)  for a more extensive
discussion of the validity and the implications of these assumptions.

\subsection{The Different Properties of Satellite and Central Galaxies}
\label{morph}
\begin{figure}
\centerline{\psfig{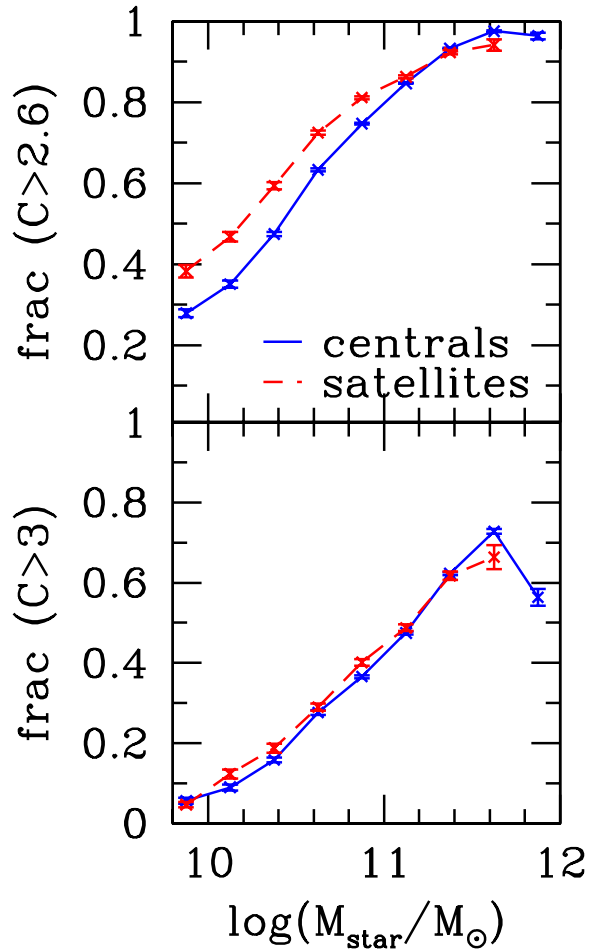}}
\caption{Top panel:  Fraction of galaxies with  $C>2.6$ for satellites
  (dashed  red line)  and central  galaxies  (solid blue  line), as  a
  function of stellar mass.  Bottom  panel: The same as above, but for
  the fraction of galaxies with  $C>3.0$. Weights have been applied in
  calculating  the  fractions. Errorbars  have  been calculated  using
  jackknife resampling.}
\label{fig:frac}
\end{figure}

In  Fig.   \ref{fig:histo},  we  show weighted  distributions  of  the
concentration,  $^{0.1}(g-r)$  colour,   $R_{50}$  and  $R_{90}$  (the
physical radius  containing 90 \%  of the Petrosian $r$-band  flux) in
four  different narrow  stellar mass  bins for  central  and satellite
galaxies.  It can  be  seen  nicely how  satellite  galaxies are  more
concentrated, redder  and smaller both  in $R_{50}$ and  $R_{90}$ than
central  galaxies, and  how  the differences  decrease towards  higher
stellar masses, where both central  and satellite galaxies are red and
concentrated.

The difference in concentration between satellite and central galaxies
is particularly interesting, as  discussed in the introduction.  Three
different  regimes  can  be   identified  in  the  three  lowest  mass
bins. Between a concentration of 1.5 and $\sim $ 2.5, the abundance of
satellite galaxies is decreased  compared to central galaxies. Between
a concentration  of $\sim$ 2.5  and $\sim $  3, it is  enhanced. Above
concentrations of  $\sim 3 $,  the abundance of central  and satellite
galaxies is the  same.  In order to further  investigate this finding,
we show in  Fig. \ref{fig:frac} the fraction of  galaxies with $C>2.6$
for centrals and satellites as a function of stellar mass (top panel),
compared  to the  fraction of  galaxies with  $C>3$ as  a  function of
stellar  mass (bottom  panel).  While  we  find that  the fraction  of
$C>2.6$ galaxies  is larger for satellites  than for centrals  up to a
stellar  mass  of  $\log(M_{\rm star}/M_{\odot})$=11,  no  significant
difference is  found for  the fraction of  galaxies with $C>3$  at all
stellar masses.  These findings have important implications. First, it
seems that  high concentration galaxies  ($C \gtrsim 3$, which  may be
identified with a population dominated by elliptical galaxies) are not
generally  produced by satellite-specific  processes. Second,  only $C
\lesssim 3$ galaxies seem to  be affected by environmental effects. It
is consistent with our findings that the concentration of a relatively
large fraction of these galaxies  is mildly enhanced after they become
satellites.  We find  no evidence  for violent  processes dramatically
altering   the   concentrations   of   satellites:  After   all,   the
environmental processes  are not strong enough  to make concentrations
reach $C>3$.

Another interesting point visible in Fig.  \ref{fig:histo} is that the
difference between  satellite and central  galaxies steadily decreases
with increasing stellar mass, and not only because galaxies in general
become more highly concentrated. In  the highest stellar mass bin, the
distribution in concentration, down  to low concentration, is the same
for  satellite  and central  galaxies.  This  may  indicate that  such
massive galaxies are not susceptible to environmental effects changing
their concentration, even if they are disk galaxies.

Motivated by  the results  described above, we  decide to  use $C_{\rm
  cut}$=3  as a  discriminator between  galaxy types.  In all  of what
follows, galaxies with $C<3$ will  be called `late types' and galaxies
with $C>3$  `early types'. Note that  in this way, our  `early type' -
`late  type'  division  also  corresponds  to a  division  into  those
galaxies  which  can be  affected  or  produced by  satellite-specific
processes,  and those which  cannot.  Using  $C_{\rm cut}=3$  has also
been suggested by Shimasaku et  al. (2001) and is slightly higher than
the cut used  for discriminating early and late  type galaxies in most
other studies.

\subsection{The stellar mass-size relation}
\label{size}

\begin{figure}
\centerline{\psfig{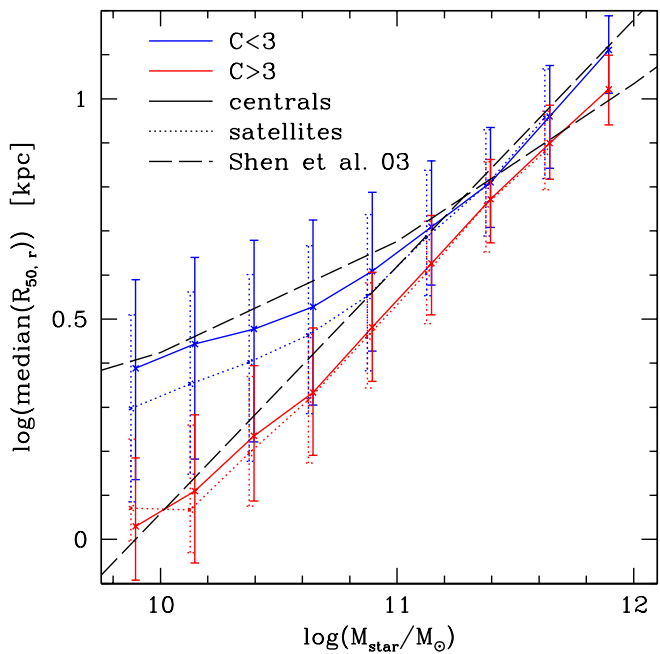}}
\caption{Logarithm of  the median  Petrosian half-light radius  in the
  $r$-band for  central (solid  lines) and satellite  galaxies (dotted
  lines). Results  for early and late  type galaxies are  shown in red
  and  blue,  respectively.   Errorbars  correspond  to  the  68  $\%$
  confidence  level  and  are  slightly  offset for  clarity  for  the
  satellite galaxies. Black lines denote  the fits obtained by Shen et
  al. (2003)  and Shen  et al. (2007),  for the  S\'{e}rsic half-light
  radii in the z-band.}
\label{fig:ser_PNm}
\end{figure}

Fig. \ref{fig:ser_PNm} shows  the median $r$-band Petrosian half-light
radius  as  a  function  of  stellar  mass for  late  and  early  type
galaxies. The sample is  additionally split into central and satellite
galaxies.  We find  that for late type galaxies,  the median radius at
fixed stellar mass is slightly smaller (by up to 20 \%) for satellites
compared to  centrals.  No  difference is found  between the  radii of
early type  centrals and satellites,  indicating that the radii  of $C
\gtrsim 3$ galaxies remain  unchanged when they become satellites. The
errorbars denote the regions where  68 \% of the measured galaxy sizes
lie for each stellar mass  bin. For comparison, the black dashed lines
show the  fits presented in  Shen et al.  (2003, 2007) for  the entire
galaxy population. While the trends  are similar, our median sizes are
clearly smaller than  those found by Shen et  al. (2003).  However, an
offset is  expected since we  have used a different  size measurement,
different stellar  mass estimates  and also a  different morphological
division.

We have carried out a series of tests to check whether our result that
late type satellites are smaller than late type centrals is robust. We
do  not  show the  results  of these  tests  here,  but describe  them
below. As a  first test, have verified that our  results do not change
qualitatively if we divide the  sample at $n=2.5$, $C=2.86$ or $C=2.6$
instead of  at $C=3$. However, we  find that the median  sizes of both
early and late type galaxies can  vary by up to $\sim$ 25 \% depending
on how  we split the  sample into early  and late types. On  the other
hand, if  we use the  S\'{e}rsic $z$-band radius  (as done by  Shen et
al. 2003)  instead of  the Petrosian $r$-band  radius, the  radii only
increase weakly,  by less  than $\sim$ 10  \%.  Additionally,  we have
also verified that  our results do not change  significantly if we use
the $z$-band or $g$-band Petrosian half-light radii.

We also  check what happens if we  include a cut in  the axis-ratio to
discriminate between  early and late  type galaxies (again,  not shown
here).  Following Maller et al.  (2008), we classify all galaxies with
$b/a<0.55$  as late  types, irrespective  of their  concentration. The
justification   for  this   is   that  such   galaxies   have  to   be
disk-like. This results  in a weak decrease (of up to  $\sim$ 7 \%) of
the median sizes of late type galaxies, but does not change our result
regarding the size differences between central and satellite late type
galaxies.  Further tests indicate that edge-on late type galaxies have
systematically  lower physical  half-light radii  compared  to face-on
late type galaxies at the same  stellar mass. This could be related to
the increased  importance of dust in  edge-on galaxies or  to the fact
that a spherical aperture is  used for a strongly elongated structure.
Another explanation  could be that stellar masses  of edge-on galaxies
might be incorrect  (although Maller et al. (2008)  claim that this is
not likely to be the case for the Bell et al. (2003) masses).

As found e.g.  by von der Linden et al.  (2007), the overestimation of
the sky background in the  SDSS images will lead to an underestimation
of  both  the  S\'{e}rsic  and  Petrosian  half-light  radii.  Guo  et
al. (2008, in prep.) claim that massive early type galaxies have their
S\'{e}rsic radii underestimated by 10 -  50 \% due to this issue. This
problem  is  present throughout  our  analysis  and  also in  Shen  et
al. (2003);  one should  therefore bear in  mind that the  real galaxy
half-light radii  are slightly higher than reported  in these studies,
especially for massive galaxies (see  also Graham \& Worley (2008) who
use effective  radii instead  of half-light radii  and find  a steeper
size-luminosity relation  for early type galaxies).  It  is known that
the problems  with sky  subtraction are more  severe in  dense regions
(e.g. von  der Linden et al.  2007), and therefore,  the smaller sizes
for the late type satellites  could in principle be a spurious effect,
coming  from the  fact that  satellites  tend to  reside in  overdense
regions.  In order  to test this, we have  repeated our analysis using
the $r$-band  half-light radii within  the 23 mag  ${\rm arcsec}^{-2}$
isophote obtained with  an improved sky subtraction by  von der Linden
et al. (2007).  We find the difference between the  sizes of late type
central and satellite galaxies to be unchanged; therefore, we conclude
that sky subtraction do not significantly affect our results.

\subsection{Average difference between satellite and centrals}
\label{diff}
\begin{figure}
\centerline{\psfig{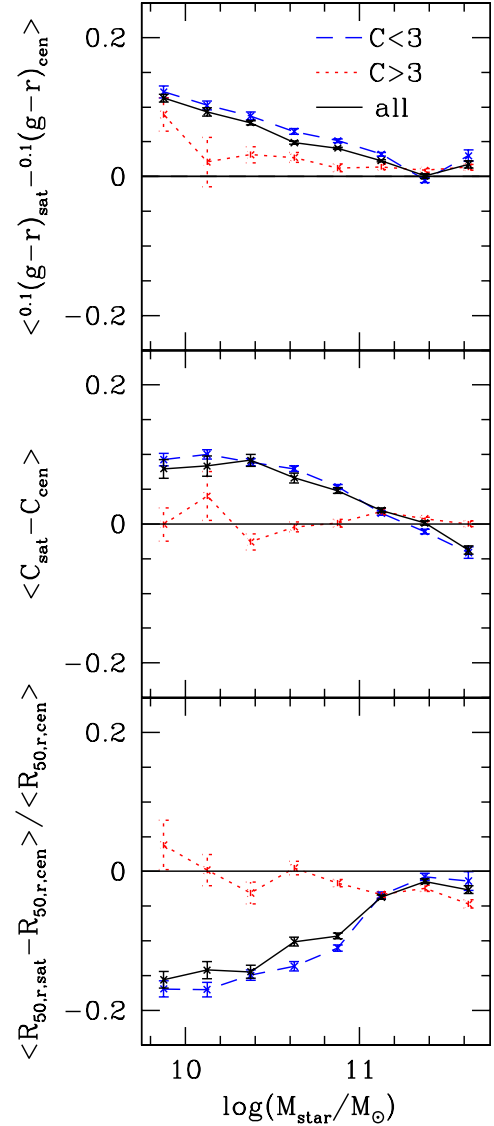}}
\caption{From  top  to bottom:  Average  difference in  $^{0.1}(g-r)$,
  concentration  and  radius  (normalized  to the  average  radius  of
  central  galaxies in each  stellar mass  bin) between  satellite and
  central galaxies  which have been  matched in stellar  mass. Results
  are  shown for  early type  galaxies  (dotted red  line), late  type
  galaxies  (dashed blue  line) and  all galaxies  (solid  black line)
  separately.   Errors   have    been   calculated   using   jackknife
  resampling. In  the bottom panel,  the error on the  average radius,
  with  which we  normalize,  has  not been  taken  into account  when
  calculating the errorbars.}
\label{fig:diff_M}
\end{figure}

In  the previous section,  we have  shown that  the median  radius for
satellite and central  galaxies differ for late type  galaxies. We now
investigate  how  this difference  in  $R_{50}$  is  connected to  the
differences  in $^{0.1}(g-r)$  and $C$  between satellite  and central
galaxies.  In order to test this, we employ the same method as used in
van den Bosch et al. (2008a).   For each central galaxy in our sample,
we first  pick a  random satellite galaxy  with the same  stellar mass
($\Delta{\rm   log}(M_{\rm  star}/M_{\odot})\leq   0.05$),   and  type
(according to concentration). We then calculate the average difference
in  radius,  colour  and  concentration  for  these  central-satellite
pairs. The  weighting scheme described  in section \ref{sec:selection}
is applied here as well:  we simply give each satellite-central pair a
weight proportional to the product  of the two weights.  The result is
shown in Fig.  \ref{fig:diff_M} by the dashed (blue)  and dotted (red)
lines.   The  black   lines  show   the  average   difference  between
central-satellite pairs which  have the same stellar mass,  but do not
have to be of the same  morphological type.  In agreement with van den
Bosch et al.  (2008a), we find that colours  of satellite galaxies are
slightly redder than the colours  of central galaxies at fixed stellar
mass (top panel of  Fig. \ref{fig:diff_M}). Interestingly, this is the
case  for both early  and late  type galaxies,  with the  effect being
slightly  stronger  for  the  late  types.  For  late  type  galaxies,
satellite galaxies  have a slightly higher  concentration than central
galaxies, while  no such  effect is seen  for the early  type galaxies
(middle  panel of Fig.  \ref{fig:diff_M}). This  is in  agreement with
what we  have found  in section \ref{morph}.   In the bottom  panel of
Fig. \ref{fig:diff_M},  one can see that late  type satellite galaxies
are smaller than late type centrals at fixed stellar mass, as found in
the  previous section.   We have  checked  that these  results do  not
change  significantly if  we also  match our  galaxies in  redshift by
requiring that $\Delta z<0.01$.

\begin{figure}
\centerline{\psfig{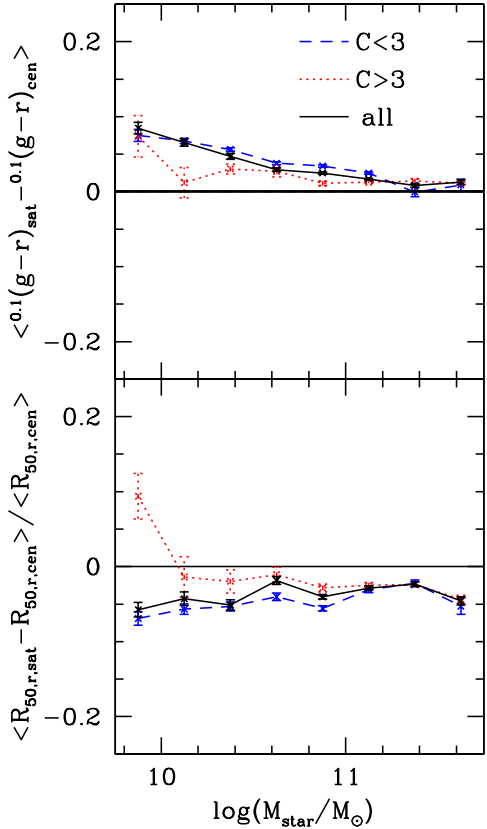}}
\caption{ As in Fig.  \ref{fig:diff_M}, but only for $^{0.1}(g-r)$ and
  radius, for galaxies matched in stellar mass and concentration.}
\label{fig:diff_MN2}
\end{figure}

\begin{figure}
\centerline{\psfig{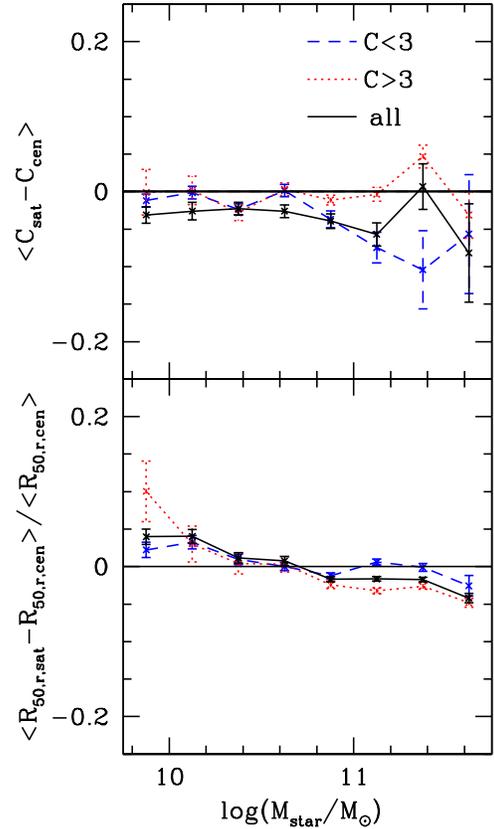}}
\caption{ As in Fig.  \ref{fig:diff_M}, but only for concentration and
  radius, for galaxies matched in stellar mass and $^{0.1}(g-r)$.}
\label{fig:diff_CM}
\end{figure}

If  we  match  galaxies  not   only  in  stellar  mass,  but  also  in
concentration  (by  additionally   requiring  that  $\Delta  C<0.05$),
differences  between  satellites and  centrals  in  colour and  radius
persist,  as   visible  in  Fig.   \ref{fig:diff_MN2}.   However,  the
difference in  radius is reduced  considerably. This indicates  that a
significant  part of  the  difference between  late  type central  and
satellite   size  is   linked   to  a   difference   in  the   average
concentration. On the other hand, if we match galaxies in stellar mass
and colour  (by additionally requiring  $\Delta ^{0.1}(g -  r)< 0.05$,
Fig.  \ref{fig:diff_CM}),  the difference in  concentration and radius
is strongly reduced  compared to if we only  match galaxies in stellar
mass, which indicates that an increase in concentration and a decrease
in size for satellites does normally not occur without an accompanying
shift towards redder colours.

To summarize,  we find that at  fixed stellar mass,  the difference in
concentration between centrals and  satellites largely arises from the
late type population.  This hints at the presence  of an environmental
process that mildly increases  the concentration in late type galaxies
which become satellites,  but does not transform late  type into early
type galaxies. We  also find that the difference  in size between late
type centrals  and satellites is  closely linked to the  difference in
concentration.  Differences  in size  and concentration, on  the other
hand, only occur with an accompanying change in colour.

\subsection{Surface Brightness Profiles for Satellites and Centrals}
\label{profile}
In the previous  two sections, we have shown  that late type satellite
galaxies are redder, smaller and have a higher concentration than late
type central galaxies of the same stellar mass.  To understand in more
detail where these differences originate, we study in this section the
surface  brightness  and  colour  profiles of  central  and  satellite
galaxies.

The  SDSS  surface brightness  profiles  are  not  often used  in  the
literature,  but have  been found  to  agree well  with deeper  images
(e.g.  Erwin,  Pohlen  \&  Beckman 2008).   The  azimuthally  averaged
surface brightness is given in a  series of up to 20 annuli with fixed
angular  sizes.  We fit  continuous  surface brightness  profiles to
these  values,  following  the  procedure  outlined  in  Stoughton  et
al.  (2002).   In order  to  be  able  to compare  surface  brightness
profiles as  a function of  physical radius for galaxies  at different
redshifts, we renormalize the surface brightness profiles, taking into
account  cosmological $(1+z)^4$  surface brightness  dimming,  so that
they all appear as at an arbitrary redshift of $z=0.1$:
\begin{equation}
\mu(R, z=0.1)=\mu(R,z)-10\log\left(\frac{1+z}{1+0.1}\right) - K_{0.1}
\end{equation}
where $\mu(R,z)$ is the  surface brightness measured in magnitudes per
square  arcseconds  at a  {\it  physical}  radius  $R$ at  a  redshift
$z$. $K_{0.1}$ is the k-correction  to $z=0.1$ according to Blanton et
al. (2003b) for  the total galaxy colour and  luminosity.  The average
profiles are calculated with the weighting scheme described in Section
\ref{sec:selection}.  We make use of the $^{0.1}(g-i)$ colour profiles
because the  difference between satellites and  centrals is especially
pronounced in this band. The  results for the $^{0.1}(g-r)$ colour are
qualitatively very similar. We also calculate the stellar mass density
profile  $\Sigma(r)$ for satellite  and central,  early and  late type
galaxies in the  two mass bins we consider,  directly from the average
colour and surface brightness profiles, using
\begin{equation}
\log[\Sigma(r)]=\varepsilon(r)-0.4  \cdot   [  \mu_{r,  0.0}(r)-m_{\rm  abs,
    \odot} ] + 8.629,
\end{equation}
where    $\Sigma(r)$    is    given   in    $M_{\odot}/{\rm    pc}^2$,
$\varepsilon(r)$=log$(M/L)_{r, 0.0}$  and $m_{\rm abs,  \odot}$ is the
absolute magnitude of the  sun in the $^{0.0}r$-band. $\varepsilon(r)$
is calculated in  each radial bin from the  average surface brightness
profiles  in  $^{0.0}r$-band  and  the  average  $^{0.0}(g-i)$  colour
profile, following the method given in Bell et al. (2003). This is the
same procedure  as used for example  in Bakos et al.  (2008).  All the
errors on the profiles are calculated using jackknife-resampling.

\begin{figure}
\centerline{\psfig{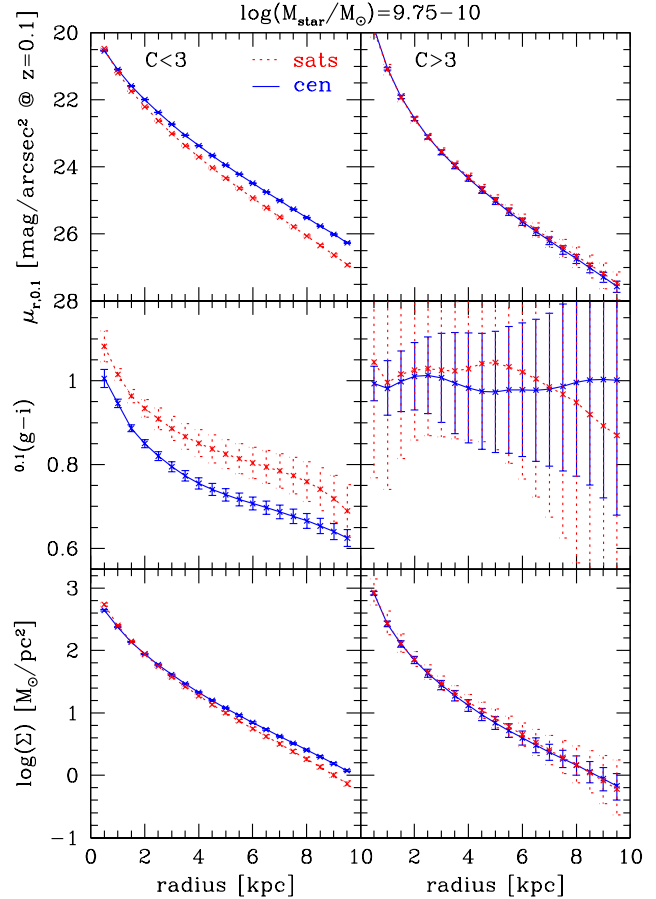}}
\caption{Top  panels:  Average  surface  brightness  profiles  in  the
  $^{0.1}r$-band for central (solid  blue line) and satellite galaxies
  (red dotted line).  Results for late type galaxies  are shown in the
  left hand panel,  results for early type galaxies  in the right hand
  panel.  Middle  panels: Average  $^{0.1}(g  -  i)$ colour  profiles.
  Bottom  panels: Stellar mass  density profiles,  directly calculated
  from the  average colour  and surface brightness  profiles.  Results
  are      shown     for      galaxies      with     $9.75<\log(M_{\rm
    star}/M_{\odot})<10.0$.  Errorbars are calculated  using jackknife
  resampling.}
\label{fig:prof_9.5MU}
\end{figure}

\begin{figure}
\centerline{\psfig{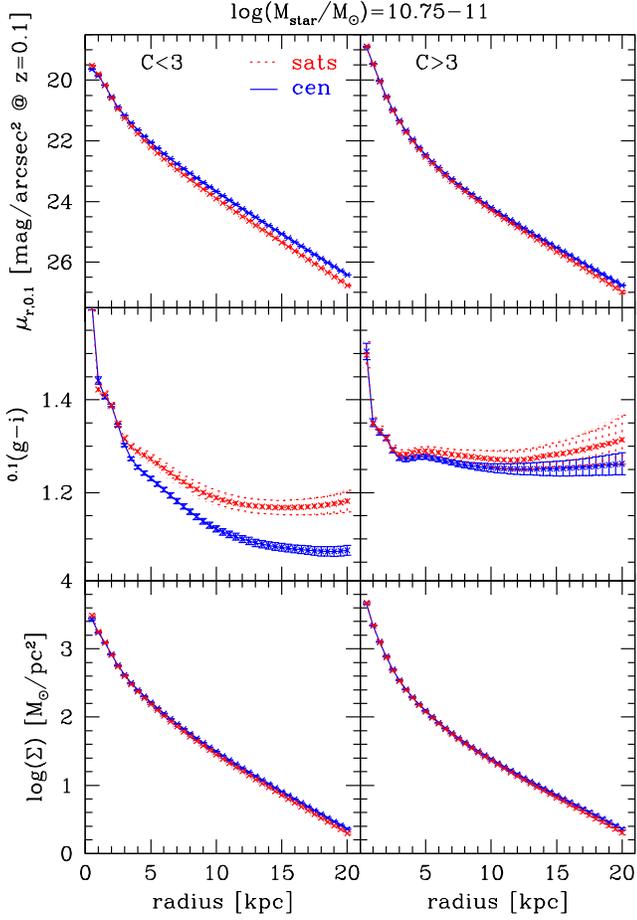}}
\caption{As in  Fig. \ref{fig:prof_9.5MU}, but for  the higher stellar
  mass bin $10.75<\log(M_{\rm star}/M_{\odot})<11.0$.}
\label{fig:prof_10.5MU}
\end{figure}

In  Fig. \ref{fig:prof_9.5MU} and  \ref{fig:prof_10.5MU}, we  show the
average  surface brightness  profile in  the top  panels,  the average
$^{0.1}(g - i)$ colour profile  in the middle panels, and the inferred
stellar mass density profiles in  the bottom panels, for galaxies with
$9.75<\log(M_{\rm    star}/M_{\odot})<10$    and    $10.75<\log(M_{\rm
  star}/M_{\odot})<11$ respectively. Results  are shown for early type
galaxies  (right  hand  panels)  and  late type  galaxies  (left  hand
panels).  In  agreement with previous  studies (e.g.  Bell \&  de Jong
2000 and  references therein; Franx, Illingworth \&  Heckman 1989), we
find that  galaxies become  bluer towards the  outskirts. For  the low
mass early  type galaxies, statistics  are not sufficient to  obtain a
reliable result.   In both stellar mass  bins, we find  that late type
satellite  galaxies have  a lower  surface brightness  than  late type
central galaxies. No difference is present in the very central region,
but it  increases towards  the outskirts of  the average galaxy  up to
$\sim$ 0.75 mag ${\rm arcsec}^{-2}$ in the lower stellar mass bin, and
up to  $\sim$ 0.4 mag ${\rm  arcsec}^{-2}$ in the  higher stellar mass
bin.  In the lower mass bin,  we find that the average colour for late
type satellite galaxies becomes  redder by a remarkably similar amount
over the entire  disk. In the higher mass  bin, the colour differences
increases towards the outskirts, but already appears at radii $\sim$ 3
kpc.   This   indicates  that  processes  which   change  colours  and
luminosities in satellite galaxies  act on (almost) the entire galaxy,
and  not only  on their  outer parts.   The fact  that the  colour and
luminosity differences  are weak  in the central  parts for  high mass
galaxies  is likely  due to  the  increasing importance  of the  bulge
component in  these systems.   Since bulges normally  have a  low star
formation rate already in isolated  galaxies, it is not to be expected
that their colours will  change strongly due to environmental effects.
For early type galaxies, we  find no difference in the average surface
brightness profile between centrals  and satellites, in agreement with
our findings that both size and concentration do not show a difference
either.   However, we  find that  the  colour profiles  of early  type
satellites  may be  redder than  for early  type centrals  by  a small
amount; this  is in agreement  with our finding that  their integrated
$^{0.1}(g-r)$ colour is slightly  redder as well (section \ref{diff}).
Generally,  the  colour differences  between  satellites and  centrals
presented here  are in good agreement  with the ones found  by van den
Bosch et  al. (2008a)  for the total  galaxy colours.   The difference
between the satellite and central stellar mass density profile (bottom
panels  of Fig.   \ref{fig:prof_9.5MU}  and \ref{fig:prof_10.5MU})  is
very  small  and  smaller  than  the difference  between  the  surface
brightness profiles, indicating  that no significant redistribution of
mass occurs  if a galaxy becomes a  satellite, as long as  it stays in
the same stellar  mass bin. The decreased stellar  mass density in the
outskirts of late type galaxies in the lower mass bin does not have to
be caused by  mass loss; rather, it is likely  that the decreased star
formation rate in satellite galaxies has the strongest impact on these
outer regions of the galaxy  which are supposed to form last (Trujillo
et al. 2006, see also discussion in the next section).

\begin{figure}
\centerline{\psfig{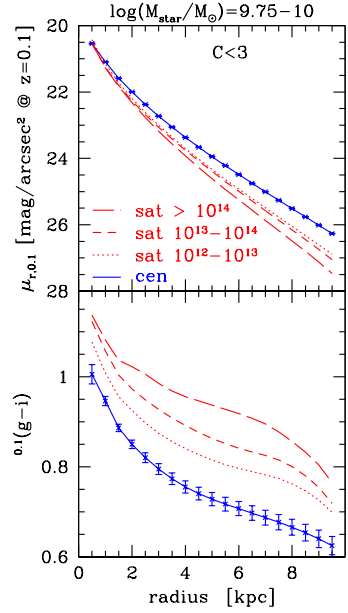}}
\caption{Top  panel:  Average   surface  brightness  profiles  in  the
  $^{0.1}r$-band for  centrals (residing in all kinds  of groups, blue
  solid lines) and satellite  galaxies (red lines), residing in groups
  with  masses in three  different bins,  as indicated.  Bottom panel:
  Average $^{0.1}(g - i)$ colour profile, as above.  Errorbars are not
  shown  for clarity.  We only  show results  for late  type galaxies,
  since   the  statistics  for   the  early   type  galaxies   is  not
  sufficient.  Results are shown  for galaxies  with $9.75<\log(M_{\rm
    star}/M_{\odot})<10$.}
\label{fig:prof_9.5EX}
\end{figure}

\begin{figure}
\centerline{\psfig{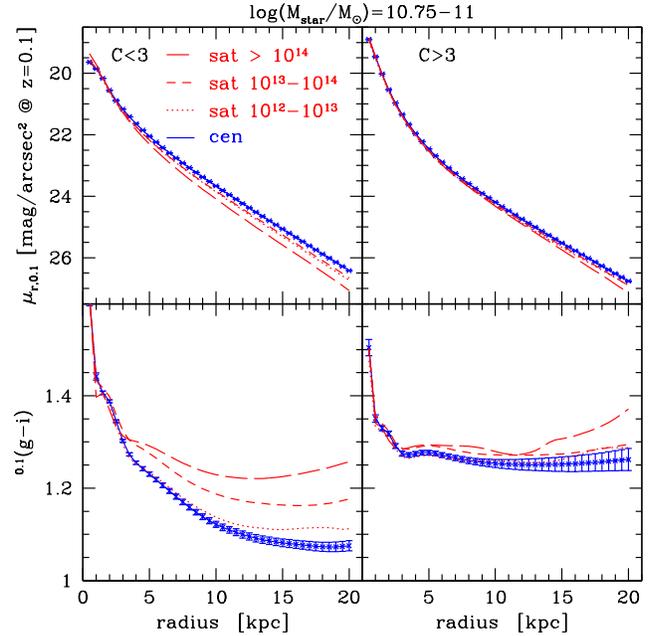}}
\caption{Same as Fig.\ref{fig:prof_9.5EX},  but in the stellar mass
  bin  $10.75<\log(M_{\rm  star}/M_{\odot})<11.0$.  Results are  shown
  both  for late  type  galaxies  (left hand  panels)  and early  type
  galaxies (right hand panels). }
\label{fig:prof_10.5EX}
\end{figure}

We  check how  our results  for  the surface  brightness profiles  and
colour profiles  change as  a function  of the host  halo mass  of the
satellites in Fig.  \ref{fig:prof_9.5EX} and \ref{fig:prof_10.5EX}. We
consider         three        bins         in        $\log\left[M_{\rm
    halo}/(h^{-1}M_{\odot})\right]$:      $12-13$,     $13-14$     and
$14-15$. Results for the early type galaxies in our lower stellar mass
bin are not shown because  of insufficient statistics.  We do not show
errorbars  either,  which  are   large  ($\sim$  0.5  for  the  colour
profiles).   We again  plot the  results for  the central  galaxies as
before  for  comparison. The  average  colour  and surface  brightness
differences  between  centrals and  satellites  clearly increase  with
increasing host halo  mass, with the change being  more pronounced for
the  galaxies in  the lower  stellar mass  bin. Interestingly,  in our
lower mass  bin, we find that  the colour change  with increasing host
halo  mass  is very  similar  across  the  entire disk  for  satellite
galaxies.   Our  results  are  in  agreement with  van  den  Bosch  et
al. (2008b,  see their Fig. 2)  who found that  the average integrated
$^{0.1}(g-r)$ colour of satellites at  fixed stellar mass is redder by
$\sim  0.05-0.1$ in  the highest  mass clusters  compared to  low mass
groups, with the  effect being stronger for galaxies  with low stellar
mass. Note that  we have found that the  integrated colour corresponds
to a colour in  the profile at around 1-4 kpc in  the two stellar mass
bins  we  consider. The  difference  in  the integrated  $^{0.1}(g-i)$
colour between  late type  satellites in the  highest and  lowest halo
mass bin is $\sim$ 0.1 in the lower stellar mass bin, and $\sim $ 0.07
in the higher stellar mass bin.  The redder colours found in satellite
galaxies  residing  in higher  mass  groups  could  indicate that  the
process  shutting down star  formation becomes  more efficient  as the
group  mass  increases.   Alternatively,   it  could  also  mean  that
satellites  which currently  reside  in clusters  have been  satellite
galaxies for a  longer time than satellites which  are currently found
in groups.   In particular, satellites currently  residing in clusters
may have been `preprocessed' as satellites in groups. Although Berrier
et al. (2008) find that most satellites in clusters have been accreted
directly from the field, the average time when a galaxy first became a
satellite might  still be different for group  and cluster satellites.
We defer this open question to future work.

It is perhaps questionable if it is useful to include edge-on galaxies
when  calculating average galaxy  profiles, since  those will  be more
strongly  affected by  dust. For  this  reason, we  have repeated  our
analysis   using   only  galaxies   with   $b/a>0.85$,  i.e.   face-on
galaxies.  Results are  not  shown  here.  We  find  that the  surface
brightness profiles and the difference between centrals and satellites
is  similar,  but  there  are  interesting deviations  in  the  colour
profiles.   The  central and  satellite  late  type  galaxies in  both
stellar mass bins are bluer by $\sim$ 0.04 over the entire disk.  This
is  likely  due to  the  fact that  inclined  late  type galaxies  are
slightly  reddened. However,  this  is  not a  large  effect, and  the
difference  between  the  colour  profiles of  central  and  satellite
galaxies remains unchanged. We  have therefore decided to show results
for the entire sample instead of only face-on galaxies.

To summarize, the main difference between a central and satellite late
type  galaxy is (i)  that the  satellite galaxy  is redder  across its
entire disk  and (ii) that it is  dimmer, and more strongly  so in its
outer regions  than in the central  regions.  It is  obvious that (ii)
must lead  to a decrease in  the half-light radius. Also,  if the disk
surrounding a central bulge is dimmed, the concentration of the system
will increase. Thus,  it is likely that the dimming  we observe is one
of the main  causes of the increase in  concentration and the decrease
in radius of late type satellites compared to late type centrals.

\section{Constraining star formation histories for satellite galaxies}
\label{model}
In this section, we will  examine whether we can place any constraints
on  the mechanisms  that suppress  star formation  in  satellites.  We
construct simple  models to explain  the changes in the  radial colour
profiles  presented in  the previous  section.  One  of our  models is
motivated by  the model  of Wang et  al.  (2007),  the other one  is a
simple truncation model, as explained below.

\subsection{Models of Star Formation in Satellites and Centrals}
Wang et al. (2007) modelled  the star formation rate (SFR) of galaxies
using the equations  below, where $t_{0}$ is the  time (measured since
the  Big  Bang) when  the  galaxy  formed,  $t_{\rm central}$  is  the
timespan   that   a   given   satellite   galaxy  has   spent   as   a
central. $\tau_{c}$  and $\tau_{s}$ characterize the  timescale of the
decline in star formation rate for centrals and satellites. The SFR as
a function of time $t$ (since Big Bang) is given by
\begin{equation}
{\rm SFR}(t)=e^{-(t-t_{0})/\tau_{c}}
\label{sfr_1}
\end{equation}
for the central galaxies and
\begin{equation}
{\rm      SFR}(t)=e^{-t_{\rm     central}/\tau_{c}}e^{-(t-t_{0}-t_{\rm
    central})/\tau_{s}}
\label{sfr_2}
\end{equation}
for the satellite galaxies,  for which, by definition, $t-t_{0}>t_{\rm
  central}$.   Using  the  abundances  and projected  correlations  of
galaxies in the SDSS as a function of 4000-$\AA$ break strengths, Wang
et al.  (2007) obtained  the characteristic timescales  $\tau_{c}$ and
$\tau_{s}$ in four different mass  bins. They find that while $\tau_c$
is a strong function of  stellar mass for central galaxies (going from
$\sim$ 40  Gyr in  low mass galaxies  to $\sim$  2.5 Gyr in  high mass
galaxies, within  the typical mass  range in the SDSS),  $\tau_{s}$ is
remarkably similar  in all  mass bins, with  values between 2.1  - 2.5
Gyr.  Such  a  slow  decline  in  SFR in  satellites  is  expected  if
strangulation/starvation  is taking  place. In  what follows,  we will
refer to this scenario as the ``slow decline'' scenario.

If the ISM is stripped by ram-pressure or tidal forces, satellites may
experience  a  very  quick  and  complete  truncation  of  their  star
formation.  As an alternative  scenario to the ``slow decline'' model,
we will therefore also check how  such a fast truncation (for which we
set $\tau_{\rm  s}=0$) would affect  the colour profiles  of satellite
galaxies  and  we  will  refer  to this  as  the  ``fast  truncation''
scenario.

The  equations  for  the  SFR  as  given above  are  defined  for  the
integrated star formation  rate in the entire galaxies.  It is however
to  be expected  that  the star  formation  history will  change as  a
function of radius in a galaxy.   The main reason for this is that the
star  formation history  is  strongly dependent  on  the surface  mass
density (Kauffmann et al. 2003; MacArthur et al. 2004; Bell \& de Jong
2000), even  more so than  on the stellar  mass. How exactly  the star
formation  history changes  as  a function  of  surface mass  density,
however, is  not clear. The fact  that regions with  higher local mass
density contain older stars and  are more metal-rich than regions with
lower local mass density can be explained in two ways in terms of star
formation history, as  pointed out in MacArthur et  al. (2004). Either
denser regions  experience an earlier  onset of star  formation, which
then declines  in a similar  way as in  less dense regions.  The other
possibility  is that  star  formation  started at  a  similar time  in
regions of different density, but was more intense and declined faster
in denser  regions.  To make our  toy model as simple  as possible, we
will   assume  that  the   first  of   those  two   assumption  holds,
corresponding to an ``inside-out'' growth of the disk. Such a scenario
is for example  supported by the findings of Barden  et al. (2005) and
Trujillo et al. (2006) who studied the stellar mass-size evolution for
late type galaxies.

\begin{figure}
\centerline{\psfig{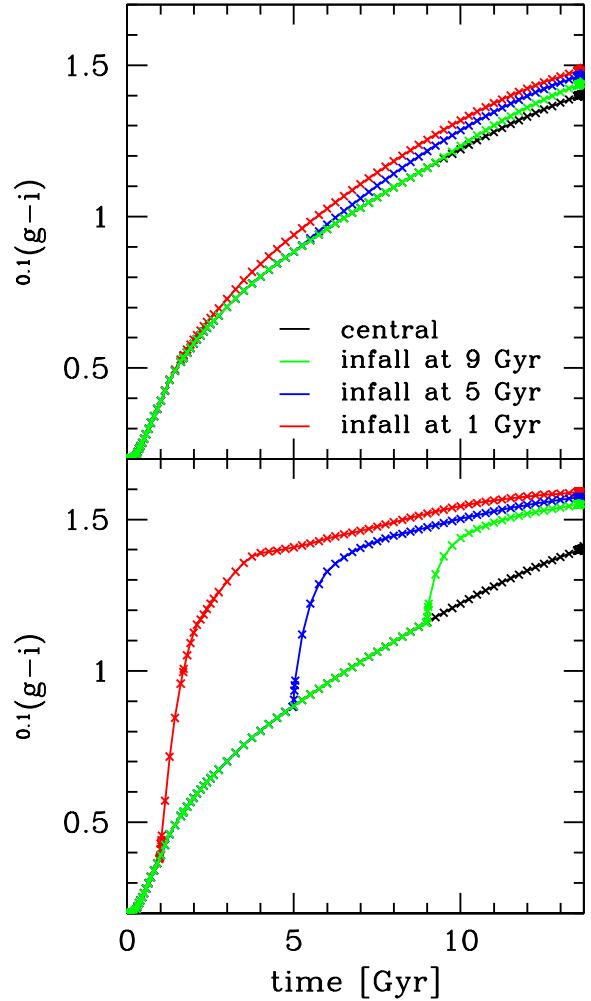}}
\caption{Evolution of  the $^{0.1}(g -  i)$ colour as function  of the
  time since onset of star formation, with the standard dust model and
  $Z$=0.02,   in  the   ``slow  decline''   (top  panel)   and  ``fast
  truncation''  (bottom panel) scenario.  Results are  shown for
  satellite galaxies with three different infall times (measured since
  onset of star formation), as  indicated, and for comparison also for
  central   galaxies,  whose   star   formation  rate   is  given   by
  eq. \ref{sfr_1}.  }
\label{fig:bc1}
\end{figure}

\subsection{Results}
\label{res_rad}
\subsubsection{Reproducing the average colour profiles}
We will  focus on galaxies with  $10.75 <\log(M_{\rm star}/M_{\odot})<
11$. For galaxies in a similar mass range, Wang et al. 2007 have found
that $\tau_c = 3.147$ Gyr  and $\tau_s=2.491$ provide the best fits to
the data.  We calculate  the $^{0.1}(g - i)$ colour\footnote{$^{0.1}(g
  - i)$ was calculated  from the $g-r$ and $g-i$  colours according to
  Blanton \& Roweis 2007.} as a  function of the lifetime of a galaxy,
using the  Bruzual \& Charlot  (2003, hereafter BC03) model,  both for
the ``slow  decline'' and the  ``fast truncation'' scenario.  The BC03
default  dust  model  is  included,   and  we  use  a  metallicity  of
$Z=0.02$. In  what follows the `formation'  of a galaxy  refers to the
start of star formation, while `infall time' is the time when a galaxy
changes from  being a central to  being a satellite.  In  the top
panel of  Fig. \ref{fig:bc1},  we show the  $^{0.1}(g-i)$ colour  as a
function of  time since formation for satellites  with three different
infall times  of 1, 5  or 9 Gyr  after formation, according  to ``slow
decline'' scenario. For comparison,  we also show the colour evolution
of central  galaxies.  In the  bottom panel of  Fig. \ref{fig:bc1},
the colour evolution for the  ``fast truncation'' model is shown. Note
that the biggest colour change occurs for galaxies which are young and
blue at the time of infall (or parts of the galaxy which are young and
blue).

We  construct  a simple  model  to  predict  the $^{0.1}(g-i)$  colour
profile  for satellite  galaxies.   We use  the average  $^{0.1}(g-i)$
colour profile for late type central galaxies with $10.75 <\log(M_{\rm
  star}/M_{\odot})< 11$ shown in Fig. \ref{fig:prof_10.5MU} and simply
assume that the  change in colour as a function of  radius is only due
to a different formation time of the stellar population dominating the
surface  brightness at  a  given radius,  assuming the  aforementioned
``inside-out'' growth  of the disk. This is  an oversimplification, as
both the metallicity and the dust  content which vary as a function of
radius   are   expected  to   contribute   to   colour  gradients   in
galaxies. However, since this is only a toy model, we do not take into
account  these  effects  here.   We  therefore assume  that  the  star
formation rate is given by eq. \ref{sfr_1} irrespective of radius, but
that $t_{0}$ varies  as a function of radius.  In  this way, we obtain
the  formation time  as a  function  of radius.  We do  not take  into
account the innermost bins. In the  first bin we consider, at a radius
of 2  kpc, we  find that the  formation time is  unrealistically high,
namely  13.3 Gyr.  This  is expected,  since  we have  not taken  into
account   the  increased   influence   of  dust,   and  the   enhanced
metallicities in the inner regions of the galaxy. In the outermost bin
at 20  kpc, we find that  the start of star  formation happened $\sim$
7.5 Gyr  ago according to our model.   As a next step,  we assume that
the formation times  as a function of radius will  be exactly the same
for a satellite galaxy and compute the colours as a function of radius
and for different infall times  for the satellite. If the satellite in
our model has already become  a satellite before the respective region
was formed in  our average central galaxy, we  simply assume that this
region does not form.

\begin{figure}
\centerline{\psfig{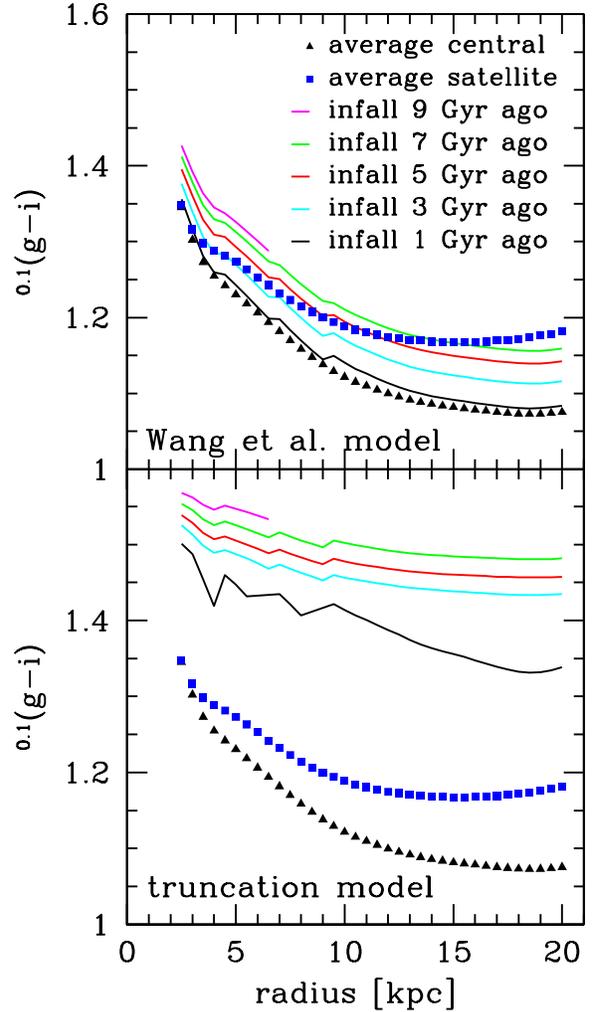}}
\caption{Black  triangles  show the  average  $^{0.1}(g  - i)$  colour
  profile of late type central  galaxies, blue squares the results for
  the colour profile  of late type satellite galaxies,  from the data,
  as  already  shown  in  Fig.  \ref{fig:prof_10.5MU},  for  $10.75  <
  \log(M_{\rm  star}/M_{\odot}) <  11$.  In  solid lines  of different
  colours, we  plot model  results for the  $^{0.1}(g - i)$  colour of
  satellite galaxies, if  we use the formation times  as a function of
  radius  as obtained  from the  central galaxies,  and  for different
  infall times for the satellite galaxy as follows: Magenta denotes an
  infall time of 9 Gyr in the past, green 7 Gyr, red 5 Gyr, cyan 3 Gyr
  and black 1 Gyr. In the upper panel, model satellite colour profiles
  are  computed according  to the  ``slow decline''  scenario;  in the
  lower panel, according to the ``fast decline'' scenario.}
\label{fig:age}
\end{figure}

The colour profiles shown from  our simple model, assuming the Wang et
al. (2007)  star formation  rates for satellites  and centrals  in the
corresponding  mass  bin,   are  shown  in  the  top   panel  of  Fig.
\ref{fig:age}. Blue squares (black  triangles) show the colour profile
of an  average late type  satellite (central) galaxy in  the according
stellar  mass bin  (without constraints  on  the host  halo mass),  as
already  shown  in Fig.   \ref{fig:prof_10.5MU}.   While a  relatively
recent infall 3 Gyr  ago seems to fit well out to  a radius of 10 kpc,
results become more unclear in  the outer regions, where the satellite
galaxies appear  to be too red  compared to the models.  This could be
explained if we assume a too  recent formation for the outer region of
the average galaxy, maybe because  our adopted metallicity is too high
for this part of the galaxy.  Another option is that a fraction of the
more loosely  bound cold gas in  the outskirts of the  galaxy is lost,
speeding  up the  decrease in  star formation  rate slightly. However,
note also that the average profiles at radii above 15 kpc are rather
uncertain. In particular, the average satellite colour profile does
not show an upturn towards redder colour anymore if we decide to remove
datapoints with errors larger than 10 \% of the total flux (not shown
here). If the
effects  of contamination  of the  satellite and  central  sample are
accounted  for  (see  section   \ref{complete}),  the  colour  of  the
satellite  galaxies becomes  redder by  up  to 0.05  in the  outermost
radial bin,  while the effect is  weaker in the  center. However, this
increases the difference between the model and data only marginally.

If  we assume  that ram-pressure  stripping of  the entire  ISM, which
leads to a  very abrupt truncation of star  formation, is the dominant
effect,  the picture  looks  very  different. In  the  bottom panel  of
Fig.  \ref{fig:age},  we perform  the  same  exercise  as before,  but
assuming  that star  formation  is completely  truncated  if a  galaxy
becomes a satellite. Clearly, this  leads to colour profiles which are
much  too  red.  In the  outermost  bin,  even  a truncation  of  star
formation only 1  Gyr before the present time leads  to $^{0.1}(g - i)
\sim$ 1.35, which  is too red by 0.15.  Of course,  one can argue that
ram-pressure  stripping of  the  ISM  will not  be  efficient for  all
satellite galaxies.  However, using the Millennium simulation combined
with analytical  estimates, Br\"{u}ggen \&  De Lucia (2008)  find that
half of the  satellites residing in haloes with  masses above $10^{14}
M_{\odot}$  today  must   have  experienced  significant  ram-pressure
stripping and they  claim that a ram-pressure of  this magnitude would
strip a galaxy of a significant amount of its cold disk gas.  Roediger
\& Hensler (2005) claim that a galaxy with a mass of $8 \times 10^{10}
M_{\odot}$ loses all its  gas down to a radius of 9  kpc if subject to
such ram-pressure for  400 Myr. A strong shift  towards redder colours
at radii above 9 kpc, as expected in such a scenario, would be visible
in our profiles, which extend to 20  kpc, but is not found. Even if we
compare  the model  results with  our  sample of  late type  satellite
galaxies   residing   in   clusters   with   masses   above   $10^{14}
h^{-1}M_{\odot}$, colours  in the ram-pressure  stripping scenario are
still  much too  red. We  therefore find  no evidence  for  the common
occurrence of such severe gas loss  in the disks of late type galaxies
in clusters, at least  out to a distance of 20 kpc  from the center of
the galaxy.

One could argue  that fast star formation truncation will
cause  galaxies to  transform  from  late type  into  early type  (for
example, if  they become a S0  galaxy) which would cause  them to fall
out of  our late type sample.  We cannot exclude that  this happens in
some cases, however, it seems that  the majority of S0 galaxies is not
formed in  this way  since their $B/T$  and luminosities do  not match
those of their potential  progenitor late type galaxies (e.g. Burstein
et al. 2005).
  
In summary,  we find  that while  the slow decline  scenario is  not a
perfect  fit to  the data  in our  simple model,   fast truncation of
star formation in a large fraction even of cluster satellite galaxies
seems much more unlikely.

\subsubsection{A potential `fast truncation' subpopulation?}
\label{subpop}
Clearly,  by only  considering the  {\it average}  colour  profiles of
galaxies, we  cannot rule  out the presence  of a small  population of
late-type galaxies with colour profiles that are in agreement with the
`fast  truncation' scenario.  In  order to  identify such  a potential
subpopulation, we look for late type galaxies with profiles similar to
those of the average early type  in our stellar mass bin. This is less
constraining  than using  the  model `fast  truncation' profiles,  and
makes  sure that we  do not  miss any  galaxies due  to the  fact that
stellar population synthesis models  tend to produce colours which are
slightly  too red,  as can  be seen  from the  comparison  between the
profiles  of  early type  galaxies  and  the  model `fast  truncation'
profiles.  As visible in Fig. \ref{fig:prof_10.5MU}, the average early
type galaxy is  not only redder than the average  late type galaxy, it
also has  a remarkably flat colour  profile at radii above  5 kpc.  We
therefore  select   late  type  galaxies  with   colours  redder  than
$^{0.1}(g-i)=1.2$ both at  5 and 10 kpc, and  with a colour difference
between these  two points  of less than  0.03.  In general,  late type
galaxies  with  red  colours  are  often  strongly  affected  by  dust
(e.g. Gallazzi et  al. 2008). However, since the  amount of extinction
should decrease  with radius (e.g. Boissier  et al. 2004),  we hope to
remove a large  part of the dusty, but  normally star forming galaxies
by our criterion that the colour profile should be flat.  We find that
the late type galaxies with early type colour profiles have an average
concentration of  $ C \sim  2.7$, compared to  $ C \sim 2.55$  for the
entire  population  of late  type  galaxies  in  this mass  bin.   The
increased concentration  for the subpopulation of  galaxies with flat,
red profiles is expected if their  disk has undergone dimming due to a
decline in star  formation. Note that due to  their high concentration
and red colours, such galaxies have likely often been misclassified as
early types in  previous studies. They likely correspond  to a part of
the visually  classified red spiral  galaxy population which  has been
identified by Bamford et al.  (2008).  We find that late type galaxies
with early type colour profiles make  up 8 \% of the central late type
population and 11 \% of the  satellite late type population.  14 \% of
the satellite in clusters with masses above $10^{14} M_{\odot}$ belong
to this  category. The  fact that late  type galaxies with  early type
colour profiles  are also found among central  galaxies and satellites
in  low mass  groups  indicates  that these  galaxies  are not  mainly
produced     by     environmental     effects    like     ram-pressure
stripping.  However,  it  is  well  possible that  there  is  a  small
subpopulation which is produced in this way.

If we  remove the  late type  galaxies with  early  type colour
profiles  from our sample,  the difference  in colour  profile between
satellite  and central  galaxies remains  practically  unchanged. This
indicates that  the difference in  colour between the  average central
and   satellite  populations  is   not  mainly   driven  by   a  small
subpopulation undergoing a big change in colour, while the rest of the
galaxies remain  unaffected.

\section{Discussion}
\label{Discussion}

\subsection{Environmental Dependence of Galaxy Radii} 
We find that late type  satellite galaxies have slightly smaller radii
than their  central counterparts at  fixed stellar mass.   Our results
fit well  with a simple picture  in which the outer  disks of galaxies
fade after they become satellites.  At fixed concentration and stellar
mass,  the difference  in radius  between centrals  and  satellites is
relatively  small, showing  that a  large  part of  the difference  in
half-light radius is linked  to a difference in average concentration.
As $C=R_{90}/R_{50}$,  this indicates that $R_{50}$  is decreased more
strongly in satellites than $R_{90}$.  We find that this is indeed the
case;  while  the  median  $R_{50}$  decreases  by up  to  20  \%  for
satellites  at fixed stellar  mass, the  median $R_{90}$  decreases by
less than 15  \%. At fixed colour and stellar mass,  we find almost no
difference in radius (and concentration) between satellite and central
galaxies.  This indicates that  the change in radius and concentration
is  intimately  connected  to  changes  in  star  formation  rate.  No
differences  in radius  is observed  between early  type  centrals and
satellites.

\subsection{Constraining Environmental Processes}
\label{constraints}
We  have found  that an  environmental process  which shuts  down star
formation in  satellites over  a timescale  of a few  Gyr is  needed in
order  to explain the  difference in  average colour  profiles between
satellites and  centrals at fixed  stellar mass. This is  in agreement
with  previous  findings  by   Kauffmann  et  al.  (2003),  Balogh  et
al. (2004b), Wang  et al. (2007) and Finn et al.  (2008).  As shown by
van  den Bosch  et  al.  (2008a) and  Skibba  (2008) with  independent
methods, this  environmental process should act in  dark matter haloes
of  all masses  (at  least down  to  their resolution  limit of  $\sim
10^{12} h^{-1}M_{\odot}$).  Its efficiency is probably somewhat higher
in   high   mass   clusters   (see  Figs.   \ref{fig:prof_9.5EX}   and
\ref{fig:prof_10.5EX}).  The  only process  which seems to  meet these
requirements    is   `starvation'   (also    called   `strangulation',
e.g.  Larson, Tinsley  \&  Caldwell 1980;  Balogh,  Navarro \&  Morris
2000), which is  the removal of the hot  gaseous halo around satellite
galaxies (or,  if one  assumes cold accretion,  the truncation  of the
infall of  cold clumps onto the  galaxy) and is assumed  to take place
via  ram-pressure  stripping or  tidal  stripping.   According to  the
findings  presented in  this paper,  it  seems plausible  that such  a
process, perhaps  combined with a partial  removal of gas  in the very
outer  regions  of the  cold  gas  disk,  could entirely  explain  the
different  colour and  surface  brightness profiles  of satellite  and
centrals,  and thus also  the small  differences in  concentration and
radius  between  satellite  and  central  galaxies  at  fixed  stellar
mass. We find  at present no need for a process  which shuts down star
formation quickly or which affects the morphology directly. Therefore,
we claim that processes like harassment (Moore et al. 1996) are likely
not important. Also,  while it is consistent with  our findings that a
process like  ram-pressure stripping removes the fuel  for future star
formation (be it in the hot  halo or in the extended cold disk outside
the region significantly populated by stars), we find it unlikely that
the ISM itself  is heavily stripped in a significant fraction of 
group and cluster galaxies, as it has  been advocated by many
studies (e.g. Gunn \& Gott 1972; Quilis, Moore \& Bower 2000).

\subsection{Comparison with other studies}
\subsubsection {The effect of environment on the cold and hot gas}
Claims of the  importance of ram-pressure stripping (of  the ISM) have
been made  by many  observational studies.  For  example, Levy  et al.
(2007) find that  disk galaxies in clusters show  HI deficiency.  This
is even found in clusters where ram-pressure stripping (of the ISM) is
not expected to occur (or  in compact groups, Rasmussen et al.  2008),
which has led to the claim that ram-pressure stripping is important in
a surprisingly  large range  of environments.  However,  HI deficiency
might also  occur as a long-term  result of starvation, if  the gas in
the disk is used up in star formation.  Rasmussen et al.  (2008) point
out that this requires star  formation rates for disk galaxies to have
been significantly higher  in the past than today  (which is possible,
but still  debated, see Bell  (2008) and references therein).   It has
also  been claimed  that  the fact  that  HI disks  are displaced  and
asymmetric  in  clusters  could   be  an  indication  of  ram-pressure
stripping;  however,  such  displacements  and  asymmetries  are  also
prevalent in the field, as  discussed in Levy et al. (2007).  Finally,
it is not clear if HI deficiency is really accompanied by a deficiency
in molecular gas, which is probably harder to remove.  Several studies
(e.g.  Casoli et al.  1998)  find no CO-deficiency in spirals galaxies
in cluster cores, although those are strongly HI deficient.

Koopmann \& Kenney (2004) find that disk galaxies in the Virgo cluster
have  their  star formation  rate,  measured  by H$\alpha$,  truncated
primarily in  the outer parts of  the disks.  They claim  that this is
evidence against starvation as  the dominant process (since this would
affect star formation in the entire disk).  Bamford et al. (2007) find
that the cluster galaxies in their sample have decreased emission-line
scalelength compared  to field galaxies, indicative  of more centrally
concentrated  star   formation.   Although  we  also   find  that  the
difference in the colour profile  is more pronounced in the outer part
than  in  the  inner part  of  galaxies,  we  have  shown this  to  be
consistent with a nearly uniform  decline in star formation across the
galaxy,   if  we   assume   an   age  gradient   in   the  disk   (see
Fig.  \ref{fig:age}). This is  due to  the fact  that the  young, blue
outer regions of galaxies undergo a more pronounced colour change when
the star formation declines.  In  a recent study, Fumagalli \& Gavazzi
(2008) find  that truncation  of the star  forming disks in  the Virgo
cluster as found  by Koopman \& Kenney (2004)  only happens in extreme
cases, and  that starvation, which affects the  entire star formation,
is  a much  more common  effect  even in  such a  massive cluster,  in
agreement with our results. Nevertheless, ram-pressure stripping of at
least  part of  the cold  ISM is  likely to  occur in  dense  cores of
clusters.  For example,  Vogt  et  al. (2004)  find  examples of  disk
galaxies in the cores of clusters which are likely infalling on radial
orbits,  and show HI-deficiencies  primarily in  the leading  edges of
their cold gas disks.

Other new  findings seem to indicate that  the gas in the  hot halo is
harder  to remove  from galaxies  than  has been  assumed in  previous
models. Semi-analytical models generally  assume that the hot gas halo
of galaxies is stripped  immediately once they become satellites. This
has been shown to produce  results in disagreement with data (Weinmann
et al. 2006b; Baldry et  al. 2006), indicating that starvation is less
efficient than  assumed.  New simulations  show that stripping  of the
hot halo  gas does not  occur instantaneously. Using  SPH simulations,
McCarthy et al. (2008) find that with typical orbital parameters, 20 -
40 \%  of the  hot gas halo  around galaxies  might still be  in place
after 10 Gyr  of orbiting in a cluster.  These simulation results have
recently been incorporated  in a new semi-analytical model  by Font et
al. (2008), with the result that colours of satellite galaxies are now
well  reproduced (see  also Kang  \&  van den  Bosch 2008).   Jeltema,
Binder \&  Mulcheay (2008)  find that a  large fraction of  early type
galaxies   in  dense  environments   have  extended   X-ray  emission,
indicating that they retain a significant amount of hot gas.

\subsubsection {The effect of environment on morphology}
A  number of recent  studies are  in agreement  with our  finding that
violent satellite-specific environmental processes transforming galaxy
morphology are likely not important and do not produce $C>3$ galaxies.
Note that we do not count major mergers, which are expected to produce
elliptical  galaxies (e.g.   Toomre \&  Toomre 1972)  as environmental
effects here, since they are part  of the formation history of a given
galaxy  and are  reflected  in  its stellar  mass.   Kauffmann et  al.
(2004), Ball, Loveday \& Brunner  (2008), van den Bosch et al. (2008a)
and  Bamford  et  al.   (2008)  all find  that  the  relation  between
environment and star formation rate or colour is more fundamental than
the  relation between  environment and  morphological  indicators like
S\'{e}rsic index or concentration.  This supports our finding that the
change in  concentration with environment is only  a secondary effect,
caused  by  decreasing  star  formation  rates.  Van  der  Wel  (2008)
distinguishes between  `structure' and `morphology'  - while structure
is  closely related  to  the bulge-to-total  luminosity  of a  galaxy,
morphology also  takes into account  features of star  formation, like
the  presence of  spiral arms.   They  find that  while morphology  is
strongly  correlated  with  galaxy  density  at  fixed  stellar  mass,
structure is  only weakly dependent on environment,  in agreement with
our  results.  Another clear  indication that  environmental processes
can quench star formation  without disturbing the mass distribution in
galaxies is  the existence of `anemic spirals'  and `passive spirals',
which have  very little  or no ongoing  star formation (van  den Bergh
1979; Goto  et al. 2003).  Their abundance is  low, which could  be an
indication that  the environmental processes  affecting star formation
is indeed  slow, as suggested  in this work.  Alternatively,  it could
also mean that such galaxies are quickly transformed into S0s (but see
Burstein et al.  2005).

\subsubsection {The effect of dust}

Based on their  observations of a relatively small  sample of galaxies
in the A901/902  supercluster, Wolf et al. (2005),  Lane et al. (2007)
and Gallazzi  et al.  (2008)  speculate that environmental  effects at
intermediate  densities could  enhance dust  extinction in  late type
galaxies.   We find no  evidence for  a common  occurrence of  such an
effect.   On the  contrary, our  late type  satellites show  a clearly
decreased $z$-band  attenuation (derived from the  Balmer decrement by
Kauffmann et al.  2003) compared  to central galaxies at fixed stellar
mass.   This  effect  is  even  stronger  for  satellites  in  massive
clusters.   This is  clear evidence  that our  finding  that satellite
galaxies  have redder colours  than central  galaxies is  unrelated to
dust.  Note that a  decrease in  $z$-band attenuation  with increasing
galaxy density has  already been found by Kauffmann  et al. (2004) and
can be explained by the fact  that the dust attenuation is highest for
star-forming, gas-rich galaxies.

\section{Summary}
\label{summary}

Using the group catalogue of Yang  et al. (2007) based on the SDSS DR4
data set, we investigate the differences between satellite and central
galaxies, to constrain  satellite-specific environmental processes. We
particularly  focus  on the  differences  in  concentration and  size.
Comparing the  full distributions  in concentration for  satellite and
central galaxies, we find that  $C>3$ galaxies are likely not produced
by environmental effects, and that a large part of $C<3$ galaxies have
their concentration mildly enhanced when becoming satellites, which is
evidence against the predominance  of violent mechanisms affecting the
concentration of  satellite galaxies.  We split our  sample into early
and  late  type galaxies  according  to  concentration, using  $C_{\rm
  cut}$=3.  For  the  first  time,  we show  clearly  that  late  type
satellite galaxies  have smaller  half-light radii than  their central
counterparts at fixed stellar mass. No difference is found between the
half-light radii of early type central and satellite galaxies.

To investigate the origin of the observed differences in concentration
and  half-light radius,  we study  the surface  brightness  and colour
profiles of  central and  satellite galaxies. We  find that  late type
satellite  galaxies are dimmed  with respect  to centrals,  and become
redder. Since  these effects are  more pronounced in the  outskirts of
galaxies, this explains the decreased size and increased concentration
in  satellite galaxies.  We  find indications  that the  environmental
effect responsible for these  changes becomes more efficient in higher
mass groups.  Early type  satellite galaxies show nearly no difference
to their central counterparts, except that there is some indication of
a weak shift towards redder colour profiles.

We also  compare the  stellar mass profiles  of central  and satellite
galaxies and find them to be very similar, indicating that the stellar
mass  distribution  is  not  significantly affected  by  environmental
processes.  Using  stellar population  synthesis models, we  show that
the  observed  changes  in  the  colour profile  between  central  and
satellite late  type galaxies  can be explained  by a slow  decline in
star  formation,  and that  they  are  are  inconsistent with  a  fast
truncation.     We   conclude    that   `starvation'    (also   called
`strangulation'), i.e. the removal of the hot extended gas halo around
galaxies, perhaps combined with a  partial stripping of the very outer
regions of the  cold gas disk, is the only process  needed in order to
explain  the environmental  dependencies  we are  able  to resolve  at
present in the SDSS.

\section*{Acknowledgments}
It is a pleasure to  thank Dimitri Gadotti for interesting discussions
about galaxy morphology,  and for providing us with  his data set. We
are grateful to the referee for his constructive comments which
helped to improve the paper. We
thank Anja von der Linden  for letting us use her corrected half-light
radii and for  useful discussion, and Simon White,  Eric Bell, 
Shiyin  Shen and Christian  Thalmann for helpful  suggestions and
useful  discussion.   SW has  been  supported  by  the Swiss  National
Science Foundation (SNF).


\end{document}

%% file: macros_wkb.tex
\def\beq{\begin{equation}}
\def\eeq{\end{equation}}
\def\barray{\begin{eqnarray}}
\def\earray{\end{eqnarray}}

\def\gtsima{$\; \buildrel > \over \sim \;$}
\def\ltsima{$\; \buildrel < \over \sim \;$}
\def\prosima{$\; \buildrel \propto \over \sim \;$}
\def\gsim{\lower.7ex\hbox{\gtsima}}
\def\lsim{\lower.7ex\hbox{\ltsima}}
\def\simgt{\lower.7ex\hbox{\gtsima}}
\def\simlt{\lower.7ex\hbox{\ltsima}}
\def\simpr{\lower.7ex\hbox{\prosima}}

\newcommand{\apj}{ApJ}

\newcommand{\aj}{AJ}
\newcommand{\mnras}{MNRAS}
\newcommand{\aap}{A\&A}

\newdimen\hssize
\newdimen\hdsize
